\documentclass[twocolumn]{aastex63}
\usepackage{multirow}
\usepackage[figuresright]{rotating}
\usepackage{subfigure}
\usepackage{epic,eepic}
\usepackage{graphicx}
\usepackage{longtable}
\usepackage{float}
\usepackage{pifont}
\usepackage{gensymb}
\usepackage{multirow}
\usepackage{amsmath}
\usepackage{color}

\shorttitle{}
\shortauthors{Zhao et al.}

\begin{document}

\title{Searching for accreting compact binary systems from spectroscopy and photometry: Application to LAMOST spectra}

\correspondingauthor{Song Wang}
\email{songw@bao.ac.cn}

\author{Xinlin Zhao}
\affiliation{Key Laboratory of Optical Astronomy, National Astronomical Observatories, Chinese Academy of Sciences, Beijing 100101, China}
\affiliation{School of Astronomy and Space Sciences, University of Chinese Academy of Sciences, Beijing 100049, China}

\author{Song Wang}
\affiliation{Key Laboratory of Optical Astronomy, National Astronomical Observatories, Chinese Academy of Sciences, Beijing 100101, China}
\affiliation{Institute for Frontiers in Astronomy and Astrophysics, Beijing Normal University, Beijing 102206, China}

\author{Jifeng Liu}
\affiliation{Key Laboratory of Optical Astronomy, National Astronomical Observatories, Chinese Academy of Sciences, Beijing 100101, China}
\affiliation{Institute for Frontiers in Astronomy and Astrophysics, Beijing Normal University, Beijing 102206, China}
\affiliation{College of Astronomy and Space Sciences, University of Chinese Academy of Sciences, Beijing 100049, China}
\affiliation{New Cornerstone Science Laboratory, National Astronomical Observatories, Chinese Academy of Sciences, Beijing 100101, China}

\begin{abstract}

{Compact objects undergoing mass transfer exhibit significant (and double-peaked) $H_{\alpha}$ emission lines. Recently, new methods have been developed to identify black hole X-ray binaries (BHXBs) and calculate their systematic parameters using $H_{\alpha}$ line parameters, such as the full-width at half maximum (FWHM), equivalent width (EW), and separation of double peaks. In addition, the FWHM--EW plane from spectroscopy and the $H_{\alpha}$ color--color diagram from photometry can be used for rapid stellar classification.
We measure the $H_{\alpha}$ and $H_{\beta}$ profiles (e.g., FWHM and EW) using the LAMOST DR9 low- and medium-resolution spectra, and calculate the systematic parameters (e.g., velocity semi-amplitude of the donor star, mass ratio, inclination angle, and mass of the accretor).
A new correlation between FWHM and $K_{\rm 2}$, $K_{\rm 2} = 0.205(18)\ \rm{FWHM}$, is obtained for cataclysmic variables (CVs) in our sample. 
Both the FWHM--EW plane and the $H_{\alpha}$ color--color diagram can distinguish CVs with FWHM $\gtrsim$ 1000 km/s from Be stars and young stellar objects (YSOs) to some extent.
To improve classification accuracy and enhance the selection of compact objects, we propose a new set of idealized filters with effective widths of 30 \AA, 130 \AA, and 400 \AA \ for the narrow $H_{\alpha}$ filter, broad $H_{\alpha}$ filter, and $r$-band filter, respectively.}

\end{abstract}


\section{Introduction}
\label{intro.sec}

Searching for compact objects and accurately measuring their parameters are crucial for establishing a more complete sample and a more comprehensive mass distribution function of compact objects. 
The traditional X-ray method has identified about 340 X-ray binaries in the Milky Way \citep{2007A&A...469..807L,2024A&A...684A.124F}, including black holes (BHs) or neutron stars (NSs), and confirmed about 25 BHs by dynamically measuring their masses \citep[BlackCAT\footnote{https://www.astro.puc.cl/BlackCAT/index.php};][]{2016A&A...587A..61C}.
On the other hand, radial velocity (RV) and astrometry have been proven to be feasible methods for uncovering quiescent compact objects \citep[e.g.,][]{2023MNRAS.521.4323E,2023MNRAS.518.1057E, 2024NatAs.tmp..215W}.

\begin{figure*}
    \center
    \includegraphics[width=0.98\textwidth]{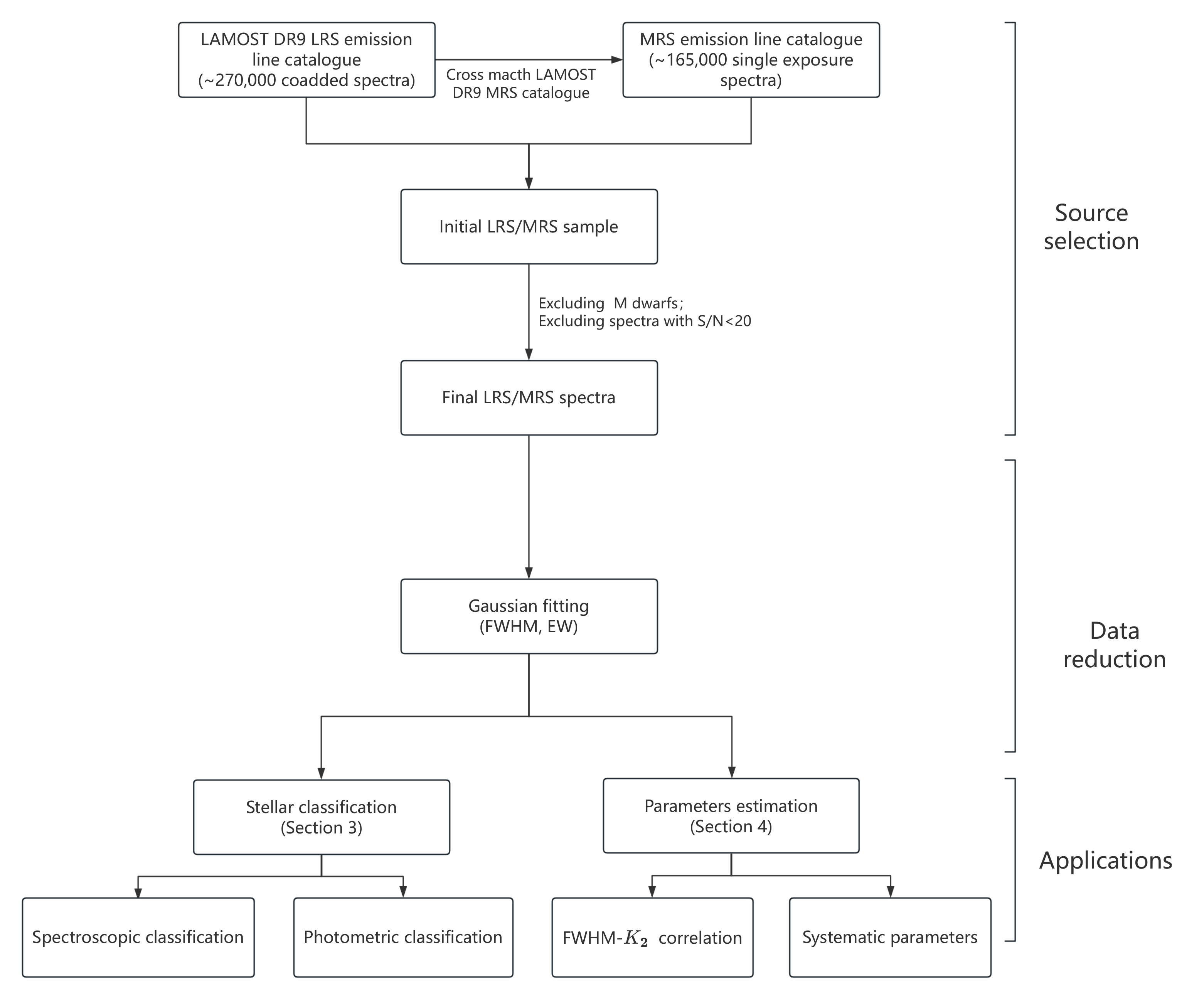}
    \caption{A summary flowchart illustrating the workflow used in this paper.}
    \label{flowchart.fig}
\end{figure*}

For X-ray binaries, the X-ray emissions originate from the accretion disks around compact objects due to ongoing mass transfer from the companion star. Measuring the masses of BHs in X-ray binaries is generally challenging. Radial velocity and light curve monitoring are essential when the system returns to its quiescent state to minimize contamination from the bright accretion disk. Recently, \citet{2015ApJ...808...80C,2016ApJ...822...99C} and \citet{2022MNRAS.516.2023C} established relations between $H_{\alpha}$ emission line parameters, such as full width at half maximum (FWHM), equivalent width (EW), and separation of double peaks, and systematic parameters including the radial velocity semi-amplitude of the donor star $K_{\rm 2}$, mass ratio $q$, and inclination angle $i$.
This provides an efficient and promising method to estimate the mass of compact objects during their accretion state.
Furthermore, \cite{2018MNRAS.481.4372C} and \citet{2018MNRAS.473.5195C} proposed a novel approach to identify black hole X-ray binaries (BHXBs) through $H_{\alpha}$ emission lines.
These authors tried to distinguish different types of objects such as BHXBs, cataclysmic variables (CVs), and Be stars in the $H_{\alpha}$ color--color diagram, which is constructed with three filters, i.e., one narrow $H_{\alpha}$ filter, one broad $H_{\alpha}$ filter, and one $r$-band filter.
These advancements make the discovery of compact objects through optical spectroscopy and photometry feasible.

Large Sky Area Multi-Object Fiber Spectroscopic Telescope (hereafter LAMOST), also known as the GuoShou-Jing Telescope, has released tens of millions of stellar spectra in low and medium resolution. It's timely to use these new methods to find compact objects in the LAMOST database and measure their systematic parameters.
As a brief introduction, LAMOST is a reflecting Schmidt telescope with an effective aperture ranging from 3.6 to 4.9 meters and a generous field of view spanning $5^{\circ}$ \citep{1996ApOpt..35.5155W}. 
LAMOST started its low-resolution spectral (LRS; $R \sim$ 1800) survey from 2011, covering a wavelength range of 3650 \AA \ to 9000 \AA.
Since October 2018, it supplemented a medium-resolution spectral (MRS; $R \sim$ 7500) survey, comprising a blue band spanning 4950 \AA \ to 5350 \AA, and a red band covering 6300 \AA \ to 6800 \AA \ \citep{2020arXiv200507210L}.

This paper aims to identify and characterise accreting compact objects by applying these new methods \citep{2015ApJ...808...80C,2016ApJ...822...99C,2022MNRAS.516.2023C} to the LAMOST DR9 sample.
This paper is organized as follows.
Section \ref{sample_data.sec} offers an overview of sample selection and data reduction processes. 
In Section \ref{color_filters.sec}, we describe stellar classification using the FWHM--EW diagram and the $H_{\alpha}$ color--color diagram, and then present a new set of idealized filters to more effectively select compact objects.
Section \ref{fwhm_ew_k2.sec} presents a new correlation between FWHM and $K_{\rm 2}$ and the estimates the systematic parameters (e.g., mass ratio, inclination, and mass of the accretor) for CVs.
Finally, a short summary is provided in Section \ref{summary.sec}.

\begin{figure*}
    \center
    \subfigure[LRS]{\includegraphics[width=0.48\textwidth]{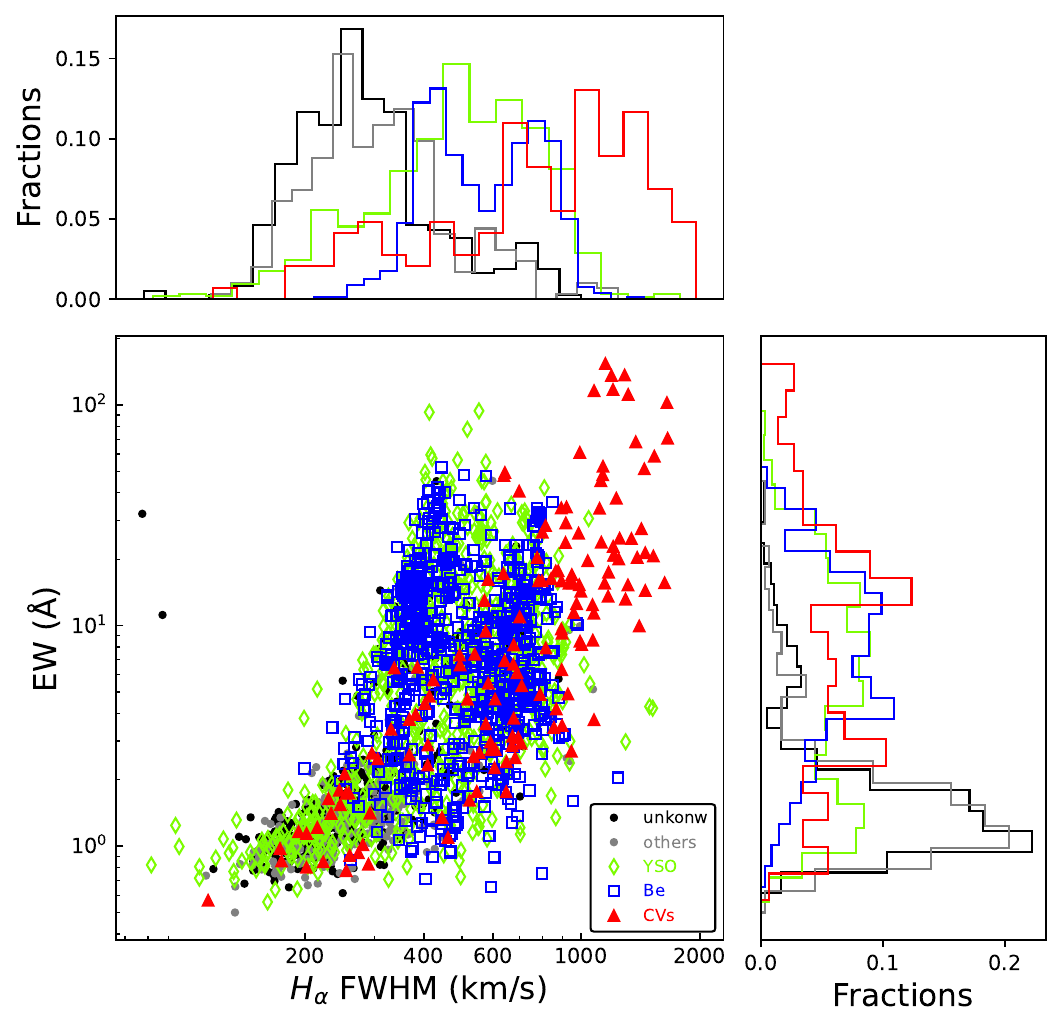}}
    \subfigure[MRS]{\includegraphics[width=0.48\textwidth]{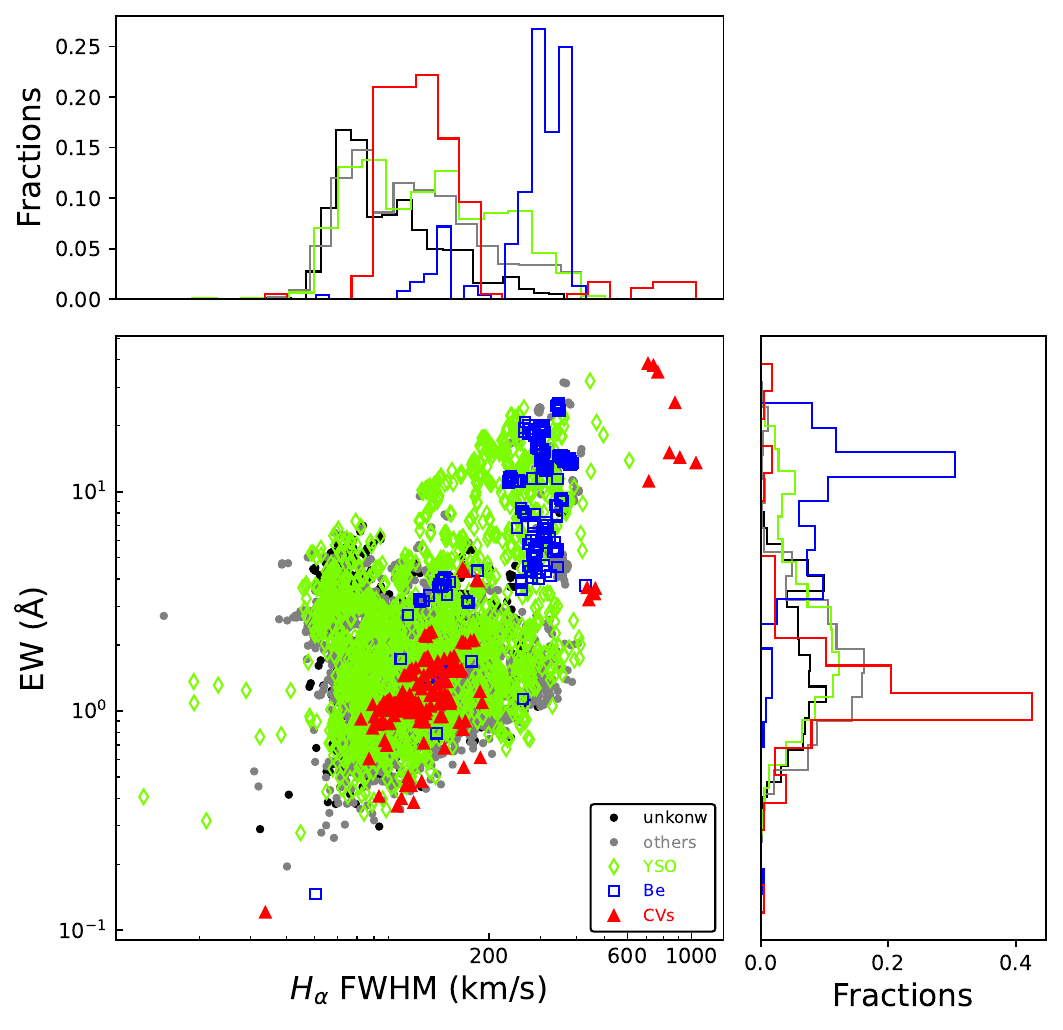}}
    \caption{FWHM--EW plane for LRS (left panel) and MRS (right panel) samples. These histograms show the fraction of population for those systems.}
    \label{fwhm_ew.fig}
\end{figure*}

\section{Sample selection and data reduction}
\label{sample_data.sec}

\subsection{Source selection}
\label{sample.sec}

Recently, Z.R. Li et al. (2024, in preparation) derived a catalog of emission-line stars using the LAMOST DR9 low-resolution spectra (LRS) data.
They employed a Pseudo Voigt function to fit more than thirty spectral lines (including the $H_{\alpha}$ line) for each spectrum.
An emission line was identified using its equivalent width (i.e., EW $>$ 0).
A total of 270,000 LRS displaying $H_{\alpha}$ emission lines were selected as our initial LRS sample.
We then cross-matched the 270,000 LRS with the LAMOST DR9 medium-resolution spectrum (MRS) catalog to build the initial MRS sample.
Furthermore, spectra with a high signal-to-noise ratio (i.e., S/N $>$ 20) were selected to obtain high quality spectral data. 
Considering that M stars often exhibit strong $H_{\alpha}$ emission lines due to magnetic activity, which could contaminate this analysis, we excluded the spectra of 7133 M stars based on the "subclass" label from the LAMOST DR9 LRS catalog, as determined by the LAMOST 1D pipeline \citep{2015RAA....15.1095L}.
We also confirmed that none of these M stars are part of the 340 X-ray binaries discovered in the Milky Way \citep{2007A&A...469..807L,2024A&A...684A.124F}, thus excluding them is a safe and necessary step.
Finally, there are 2583 LRS for 1924 stars and 6002 MRS for 771 stars in our final sample.

\subsection{Data reduction}
\label{data.sec}

Following previous works \citep{2015ApJ...808...80C,2016ApJ...822...99C},
we used a Gaussian function to fit the $H_{\alpha}$ emission lines (with RV correction), with $\sigma$ set ranging from 0 \AA\ to 50 \AA, to estimate their FWHM and EW values.
A summary flowchart outlining the source selection and data reduction was shown in Figure \ref{flowchart.fig}.

One should note that the FWHM measurements vary between different observations due to differing instrumental broadening and sampling intervals of wavelength at different resolutions.
Therefore, additional corrections to the actual measurements are necessary for the LRS and MRS samples.
Here we synthesized a group of Gaussian profiles with different $\sigma$ values and resampled the profiles to match the resolutions of the LRS and MRS.
The FWHM$_{\rm mod}$ for the original spectra were calculated as FWHM$_{\rm mod}$ $=$ 2.355$\sigma$.
The FWHM$_{\rm fit}$ for the LRS and MRS resolution were estimated with Gaussian fitting.
In the following analysis, the FWHM measurements for the LRS and MRS were corrected by adding the difference between FWHM$_{\rm fit}$ and FWHM$_{\rm mod}$ shown in Figure \ref{fwhm_correction.fig} (Appendix \ref{fwhmm_fwhmg_single_profiles.sec}).

\begin{figure*}
    \center
    \includegraphics[width=0.98\textwidth]{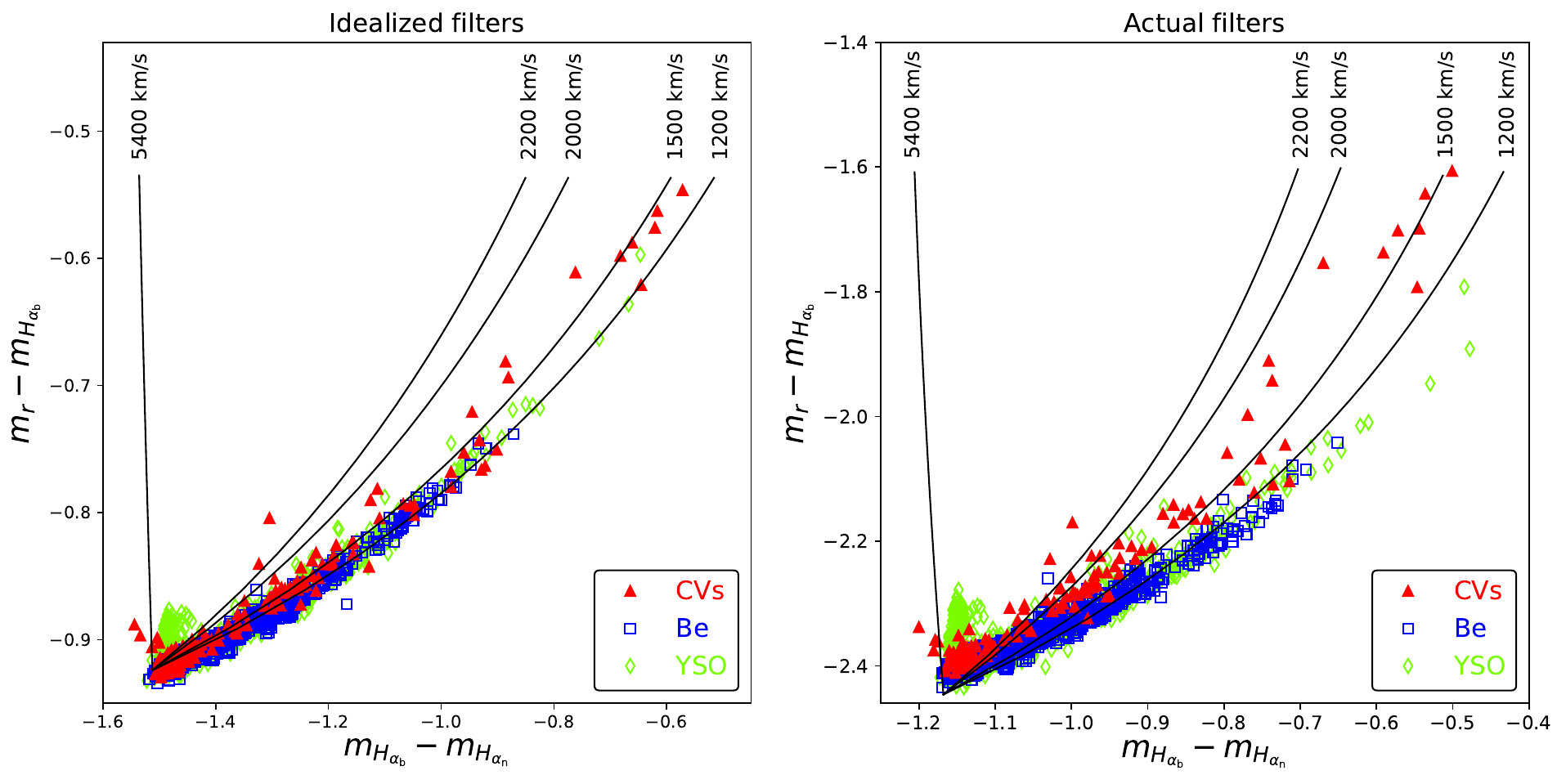}
    \caption{Left panel: $H_{\alpha}$ color--color diagram, constructed with idealized filters, using the LRS sample. The black lines stand for the results of the synthetic $H_{\alpha}$ profiles with a constant FWHM from 1200 km/s to 5400 km/s. 
    Right panel: $H_{\alpha}$ color--color diagram, constructed with actual filters, using the LRS sample. The idealized filters are centered at 6563 \AA \ with effective widths of 37 \AA, 150 \AA, and 350 \AA, while the actual filters correspond to the NOT21 narrow $H_{\alpha}$ filter, the NOT29 broad $H_{\alpha}$ filter, and the $r$-band filter from OASIS.}
    \label{LRS_bn_rb.fig}
\end{figure*}

\subsection{Stellar types}
\label{types.sec}

For the following analysis, which includes systematic parameter estimation and stellar classification using spectra and photometry, we first collected stellar types for the sample objects from previous studies.
We cross-matched our sample with different catalogs for XBs \citep{2005A&A...442.1135L,2006IAUS..230...41L}, CVs \citep{2018RAA....18...68H,2020AJ....159...43H,2021ApJS..257...65S,2023AJ....165..148H}, Be stars \citep{2016RAA....16..138H,2022ApJS..259...38Z}, YSOs \citep{2016MNRAS.458.3479M,2018A&A...619A.106G,2019MNRAS.487.2522M,2021ApJS..254...33K,2023A&A...674A..21M,2023A&A...674A..14R}, and SIMBAD.
The types from the catalogs were preferred. When one object was classified into different types by these catalogs, if it was archived in SIMBAD, the type from SIBMAD (i.e., the ``main\_type" label) was used; otherwise, we classified it in the order: CVs $>$ Be stars $>$ YSOs. 
Simultaneously, objects classified as QSO, AGN, and GALAXY by SIMBAD were removed.
Unfortunately, no known XBs were found in our sample.
Finally, the sample can be classified into five types: CVs, Be stars, YSOs, ``others", and ``unknown". The ``others" type refers to various stellar objects from SIMBAD classification, while the ``unknown" type means no classification was derived.

\section{Stellar classification}
\label{color_filters.sec}

\subsection{The FWHM--EW distribution}
\label{FWHM_EW.sec}

As shown in \citet{2015ApJ...808...80C}, objects with $H_{\alpha}$ emission, such as XBs, CVs, Be stars, and YSOs, exhibit different FWHM and EW values.
To identify possible BHXBs and CVs in the LAMOST data, we investigated the distributions of FWHM and EW for our sample based on the stellar types from Section \ref{types.sec}.

Figure \ref{fwhm_ew.fig} shows the FWHM and EW distributions for our sample objects. The $H_{\alpha}$ emission of YSOs is related to their magnetic activities. 
It appears that the YSOs can be divided into two subpopulations: a more active sample (EW $\gtrsim$ 3 \AA) and a less active sample (EW $\lesssim$ 3 \AA).
The less active YSOs, the ``others", and ``unknown" types have similar FWHM and EW distributions, while the more active YSOs and Be stars are located in a similar region in the FWHM--EW plane.
CVs display a wide range of FWHM and EW values. The narrow $H_{\alpha}$ emission lines (e.g., small FWHM values) may originate from the magnetically active companion stars during the quiescent state of the CVs, while the broad $H_{\alpha}$ emission lines (e.g., large FWHM values) are from the accretion disk during the (violent) accretion state.
Because the accretion disks have high rotational velocities, the emission lines for CVs in which accretion disks are dominant can be very broad (often with FWHM $\gtrsim$ 1000 km/s), and thus easily distinguished from other types of systems.

\begin{figure*}
    \center
    \includegraphics[width=1\textwidth]{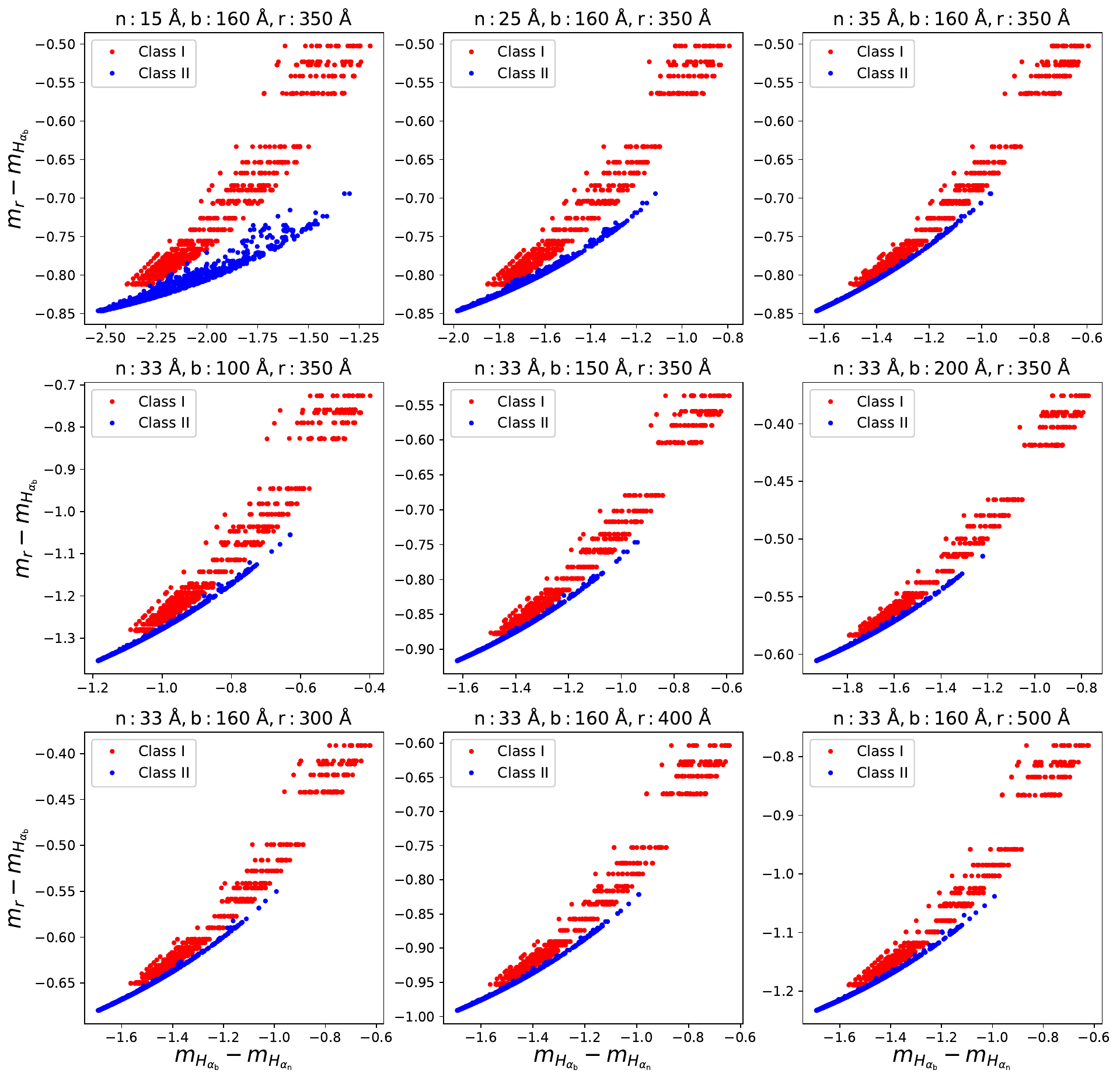}
    \caption{$H_{\alpha}$ color--color diagram in different sets of filters. The letters n, b, and $r$ in the titles refer to the narrow $H_{\alpha}$ filter, broad $H_{\alpha}$ filter, and $r$-band filter, respectively. The value following each letter represents the effective width of the filter. Class I includes CVs with FWHM $>$ 1000 km/s and EW $>$ 10 \AA, while Class II includes Be stars and YSOs with smaller FWHM and EW values.}
    \label{width_change.fig}
\end{figure*}

\begin{figure*}
    \center
    \includegraphics[width=0.98\textwidth]{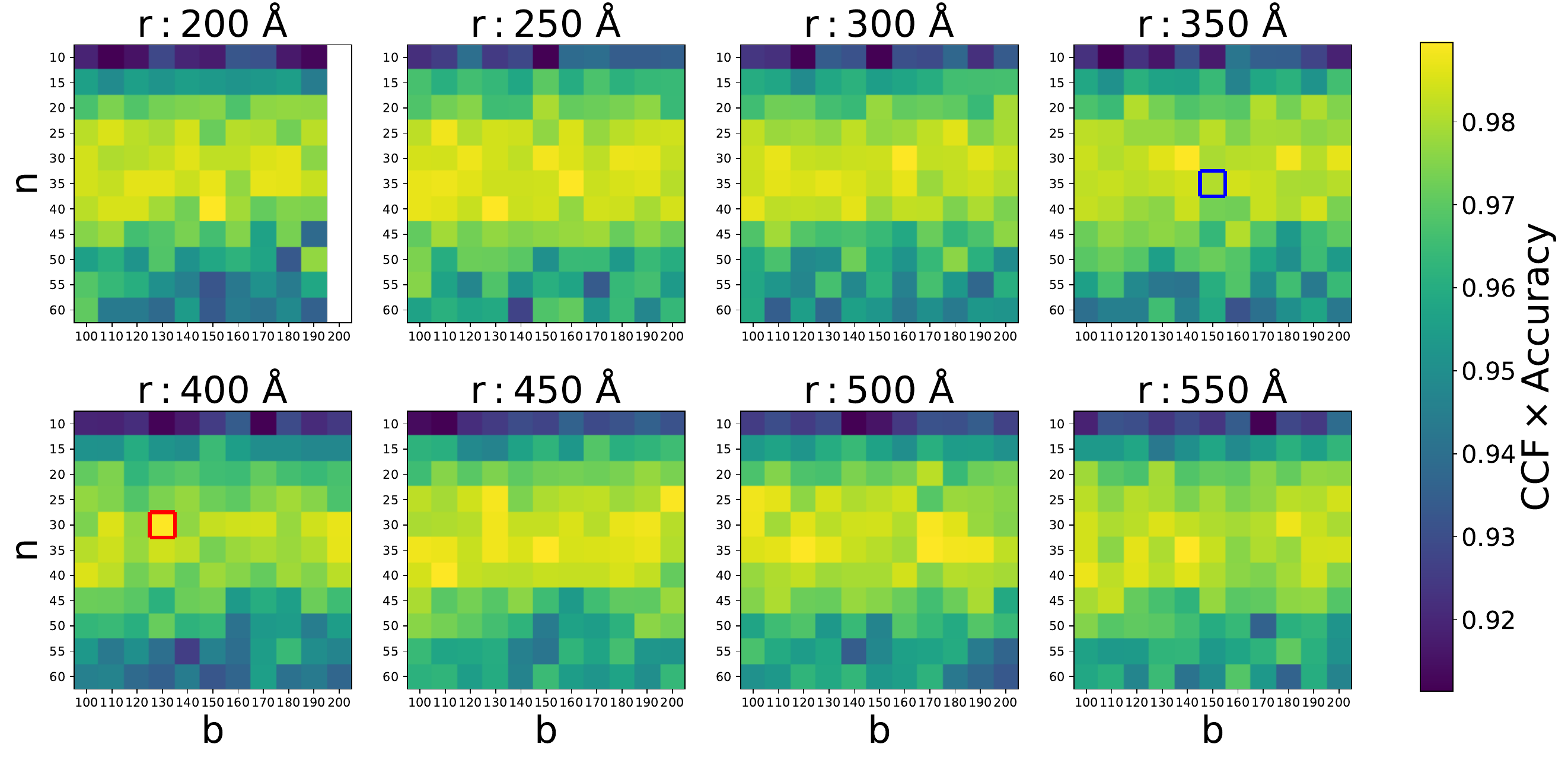}
    \caption{The accuracy of the Random Forest Classifier and the CCFs of $m_{r} - m_{H_{\alpha_{\rm b}}}$ and $m_{H_{\alpha_{\rm b}}} - m_{H_{\alpha_{\rm n}}}$ estimated at the different sets of filters. The letters n, b, and $r$ in the titles refer to the narrow $H_{\alpha}$ filter, broad $H_{\alpha}$ filter, and $r$-band filter, respectively. The red square marks the location of the best set of filters in this work, while the blue square is the location of the idealized filters used in \cite{2018MNRAS.481.4372C}.}
    \label{new_filters.fig}
\end{figure*}

\subsection{The $H_{\alpha}$ color--color diagram}
\label{color_disgram.sec}

As shown in Section \ref{FWHM_EW.sec}, the distributions of FWHM and EW can be used to distinguish different types of objects.
Besides spectra, the FWHM and EW values can also be derived from the photometric measurement of the $H_{\alpha}$ lines.
%\citep{2018MNRAS.473.5195C}.
%
\citet{2018MNRAS.481.4372C} presented a photometric method, using an $H_{\alpha}$ color--color diagram, to classify different types of objects, such as BHXBs, CVs, etc.
Those authors employed a combination of $H_{\alpha}$ and $r$-band filters, including the NOT21 narrow $H_{\alpha}$ filter\footnote{\url{http://svo2.cab.inta-csic.es/theory/fps3/index.php?id=NOT/NOT.021 \&\&mode=search\&search\_text=NOT/NOT.021\#filter}}, the NOT29 broad $H_{\alpha}$ filter\footnote{\url{http://svo2.cab.inta-csic.es/theory/fps3/index.php?id=NOT/NOT.029 \&\&mode=search\&search\_text=NOT/NOT.029}}, and the $r$-band filter MR661\footnote{\url{http://https://astro.ing.iac.es/filter/filtercurve.php?format=txt\&filter=585}} from OASIS.
They further used idealized filters (i.e., 100\% transmission) to investigate the feasibility of this method.
These idealized filters centered at 6563 \AA\ with effective widths of 37 \AA\ (narrow $H_{\alpha}$ filter), 150 \AA\ (broad $H_{\alpha}$ filter), and 350 \AA\ ($r$-band filter), respectively.

In this Section, we used LAMOST LRS data to construct the $H_{\alpha}$ color-color diagram for our sample.
Different with the original works \citep{2018MNRAS.473.5195C,2018MNRAS.481.4372C}, which performed photometry using their filters, we derived the magnitudes ($m_{r}$, $m_{H_{\alpha_{\rm b}}}$, and $m_{H_{\alpha_{\rm n}}}$) of our sample sources by convolving the LAMOST LRS data with the transmission curves of both idealized and actual filters.
The wavelength ranges of the broad $H_{\alpha}$ filter and the $r$-band filter cover the whole $H_{\alpha}$ profiles in the LRS sample, indicating that the color $m_{r}$ - $m_{H_{\alpha_{\rm b}}}$ reflects the continuum differences among different objects, while the color \( m_{H_{\alpha_{\rm b}}} - m_{H_{\alpha_{\rm n}}} \) represents the differences of the \( H_{\alpha} \) line profiles.

To include the FWHM and EW information in the $H_{\alpha}$ color--color diagram, we need to synthesize a set of double-peaked $H_{\alpha}$ profiles with different FWHM and EW values and calculate their photometry in the three bands.
Here the FWHM and EW represent observational values specific to LAMOST LRS.
Unlike EW, FWHM measurements differ between modeled and observed spectra.
Appendix \ref{fwhmm_fwhmg.sec} shows that comparison of FWHM values obtained from modeled spectra and these spectra at LAMOST/LRS resolution. Consequently, to obtain observational FWHM values of 1200 km/s, 1500 km/s, 2000 km/s, 2200 km/s, and 5400 km/s, the model FWHM values for $H_{\alpha}$ line construction should be set as 1244.51 km/s, 1562.71 km/s, 2091.46 km/s, 2302.41 km/s, and 5636.28 km/s, respectively. The EW was set within the range of 0 to 200 \AA.
The synthetic $H_{\alpha}$ profile can be created as follows \citep{2018MNRAS.473.5195C},
\begin{equation} 
f = f_{H_{\alpha}} + f_{\rm cont},
\end{equation}
where
\begin{equation} \label{eq6}
f_{H_{\alpha}} = \frac{EW}{5\sigma}\times(e^{-\frac{(\lambda-(\lambda_{0}-DP/2))^{2}}{2\sigma^{2}}}+e^{-\frac{(\lambda-(\lambda_{0}+DP/2))^{2}}{2\sigma^{2}}}) 
\end{equation}
and
\begin{equation} \label{eq7}
f_{\rm cont} = 0.0005\times(\lambda-\lambda_{0})+1.
\end{equation}
Here $f_{H_{\alpha}}$ and $f_{\rm cont}$ are the flux densities for the $H_{\alpha}$ line and continuum, respectively. 
$\lambda_{0}$ is the rest wavelength of $H_{\alpha}$, DP is defined as FWHM/1.785, and $\sigma$ is set to be DP/3.
The photometric flux $F_{\lambda}$ and magnitudes $m_{\lambda}$ for a specific filter can be calculated following
\begin{equation}\label{eq10}
F_{\lambda} = \int_{\lambda_{1}}^{\lambda_{2}} G(\lambda)f d\lambda
\end{equation}
and 
\begin{equation}\label{eq12}
\begin{split}
m_{\lambda} = -2.5{\rm log} F_{\lambda},
\end{split}
\end{equation}
respectively, where $G(\lambda)$ is the transmission curve of the filter.
$\lambda_{1}$ and $\lambda_{2}$ represent the lower and upper limits of the effective widths.

Figure \ref{LRS_bn_rb.fig} shows the $H_{\alpha}$ color--color diagram for the CVs, Be stars, and YSOs in our LRS sample using the filters in \cite{2018MNRAS.481.4372C}.
The diagram of both sets of filters (i.e., idealized and actual) shows that CVs with broad $H_{\alpha}$ lines (e.g., with FWHM $\gtrsim$ 1000 km/s), are easily distinguishable from other objects.

\subsection{The filters}
\label{filters.sec}

It is worthwhile to explore a more suitable set of filters for easier and more accurate classification of different types of objects. In the following analysis, we used idealized filters to carry out the investigation. The effective width of the filters is the only variable under consideration.

First, We divided our sample into two classes: (1) Class I, including CVs with FWHM $>$ 1000 km/s and EW $>$ 10 \AA; (2) Class II, including Be stars and YSOs with smaller FWHM and EW values.

Second, we created a series of filters with the following width ranges:
from 10 \AA \ to 60 \AA, in steps of 5 \AA, for the $H_{{\alpha}_{\rm n}}$ filter;
from 100 \AA \ to 200 \AA, in steps of 10 \AA, for the $H_{{\alpha}_{\rm b}}$ filter;
from 200 \AA \ to 550 \AA, in steps of 50 \AA, for $r$-band filter.
For each group of filters, we calculated the magnitudes ($m_{r}$, $m_{H_{\alpha_{\rm b}}}$, and $m_{H_{\alpha_{\rm n}}}$) of our sample sources by convolving their LAMOST LRS data with the transmission curves of these idealized filters.
Simultaneously, we sampled 1000 points from both Class I and Class II to balance the sizes of the two samples.
It is clear that Class I and Class II show different distributions and dispersions in the $H_{\alpha}$ color-color diagram when using different filter sets (Figure \ref{width_change.fig}).

\begin{figure}
    \center
    \includegraphics[width=0.48\textwidth]{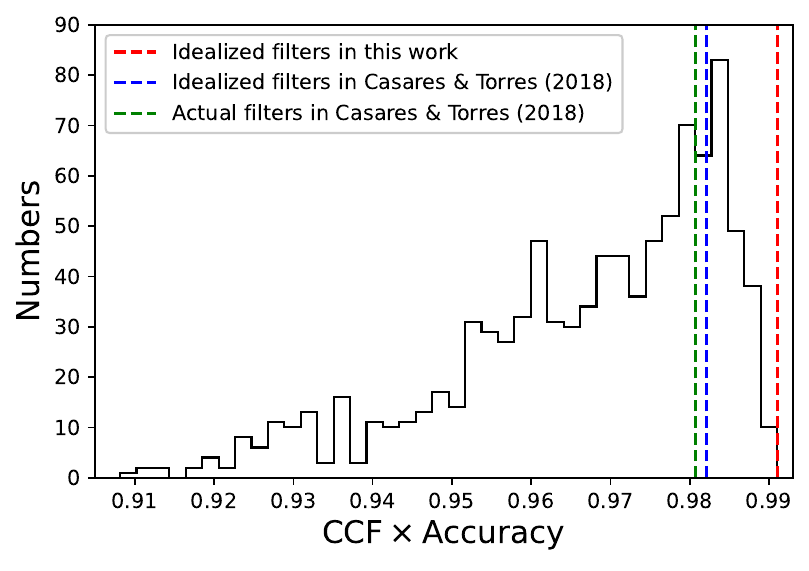}
    \caption{The histogram of the accuracy of the Random Forest Classifier and the CCF result for all filter sets. The red dash line represents the location of our idealized filters, while the blue and green dash lines are the idealized and actual filters used in \cite{2018MNRAS.481.4372C}.}
    \label{accuracy_filters.fig}
\end{figure}

\begin{figure}
    \center
    \includegraphics[width=0.48\textwidth]{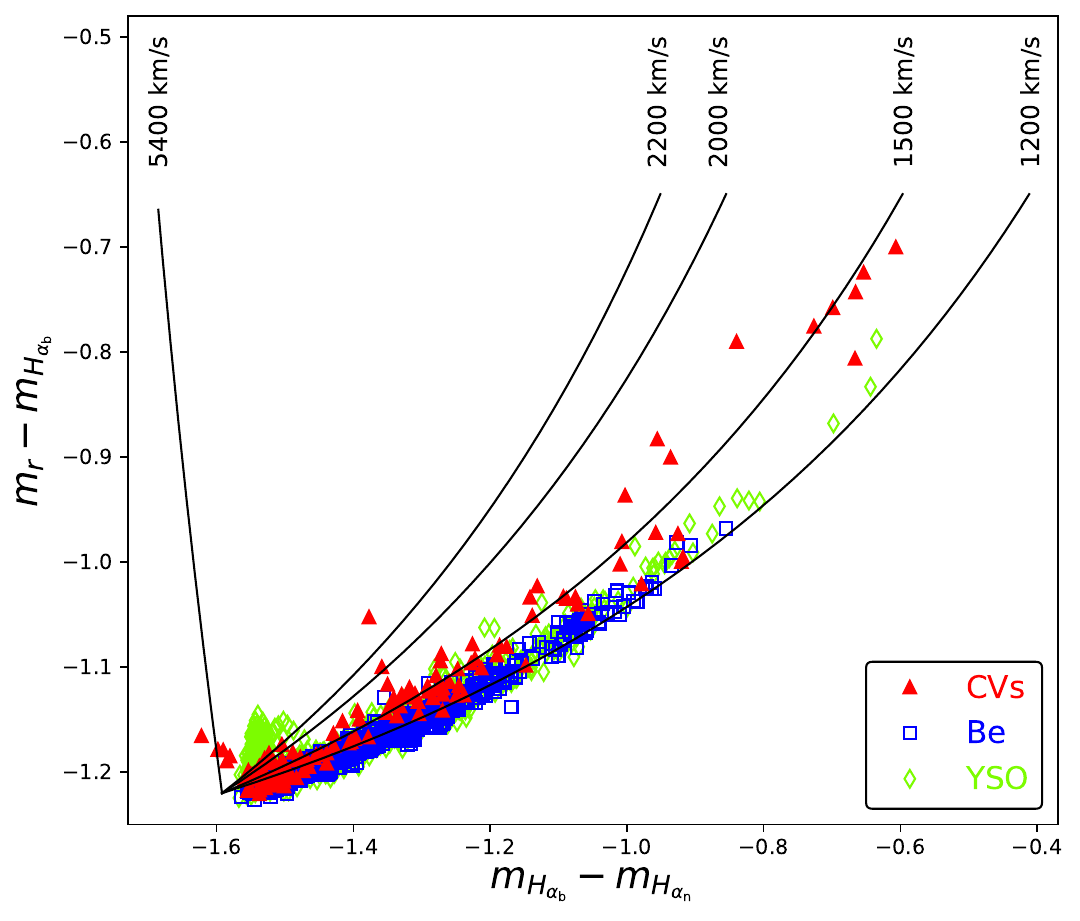}
    \caption{The $H_{\alpha}$ color--color diagram using the optimal set of filters (30 \AA\ for $H_{\alpha_{\rm n}}$ filter, 130 \AA\ for $H_{\alpha_{\rm b}}$ filter, and 400 \AA\ for $r$-band filter).
    \label{bn_rb_newfilters.fig}}
\end{figure}

Third, we tried to determine the most suitable set of filters for optimally classifying the sources in Class I and II.
On one hand, we employed a Random Forest Classifier, assigning 85\% of each class to the training set and 15\% to the test set, to differentiate Class I and II.
On the other hand, we estimated the cross-correlation function (CCF) of
$m_{r} - m_{H_{\alpha_{\rm b}}}$ and $m_{H_{\alpha_{\rm b}}} - m_{H_{\alpha_{\rm n}}}$ to quantify the dispersion of Class II.
Based on the accuracy of the Random Forest Classifier and the CCF result, we identified the optimal filter widths as 30 \AA\ for the $H_{\alpha_{\rm n}}$ filter, 130 \AA\ for the $H_{\alpha_{\rm b}}$ filter, and 400 \AA\ for the $r$-band filter (Figure \ref{new_filters.fig}).
Figure \ref{accuracy_filters.fig} shows the histogram of the accuracy of the Random Forest Classifier and the CCF result (i.e., $CCF \times Accuracy$) for all filter sets.
The $CCF \times Accuracy$ value of the optimal filter is about 0.991, compared to 0.982 and 0.981 for the ideal filter set and the actual filter set proposed by \cite{2018MNRAS.481.4372C}, respectively. This suggests that the potential improvement offered by the optimal filter sets may be limited.
Table \ref{accuracy_10syst.tab} lists the values for 10 optimal filter sets.
Figure \ref{bn_rb_newfilters.fig} shows the $H_{\alpha}$ color--color diagram using the optimal set of filters, which achieves improved separation between CVs and other objects (e.g., Be stars and YSOs) compared to the idealized filters used by \cite{2018MNRAS.481.4372C}.

\begin{table}
\caption{The $CCF \times Accuracy$ values of 10 optimal filter sets. \label{accuracy_10syst.tab}}
\centering
\begin{center}
\begin{tabular}{cccc}
\hline\noalign{\smallskip}
$H_{\alpha_{\rm n}}$ filter & $H_{\alpha_{\rm b}}$ filter &  $r$-band filter & $CCF \times Accuracy$ \\
(\AA) & (\AA) & (\AA) &   \\
\hline\noalign{\smallskip}
30 & 130 & 400 & 0.991 \\
40 & 150 & 200 & 0.991 \\
30 & 140 & 350 & 0.990 \\
35 & 140 & 550 & 0.989 \\
35 & 160 & 250 & 0.989 \\
35 & 170 & 500 & 0.989 \\
30 & 160 & 300 & 0.989 \\ 
40 & 130 & 250 & 0.989 \\
30 & 180 & 350 & 0.989 \\
35 & 120 & 500 & 0.989 \\
\noalign{\smallskip}\hline
\end{tabular}
\end{center}
\end{table}

\subsection{The theoretical distribution of CVs and BHXBs in $H_{\alpha}$ color--color diagram}
\label{colour_dis.sec}

In this Section, we aim to explore the theoretical distribution of CVs and BHXBs in the $H_{\alpha}$ color--color diagram, based on their FWHM distributions and using our optimal filter set.

By using the observed distribution of systematic parameters, \citet{2018MNRAS.473.5195C} obtained a width cut-off of FWHM $=$ 2200 km/s between the BHXBs and CVs.
Following their results, we estimated the theoretical $H_{\alpha}$ color--color diagram with the FWHM distribution ranges from 1500 km/s to 5400 km/s for BHXBs and from 1000 km/s to 2200 km/s for CVs.

Applying Equations. \ref{eq6}, \ref{eq7}, and \ref{eq10}, the theoretical photometric flux for our optimal idealized filters was calculated as follows,
\begin{equation}\label{eq11}
F_{\lambda} = \frac{EW}{2}\times(erf(\frac{W_{\lambda}+DP}{2\sqrt{2}\sigma})+erf(\frac{W_{\lambda}-DP}{2\sqrt{2}\sigma})) + W_{\lambda},
\end{equation}
where the $erf(x)$ is the error function and $W_{\lambda}$ is the effective width (i.e., $\lambda_{2} - \lambda_{1}$ in Eq. \ref{eq10}).

\begin{figure}
    \center
    \includegraphics[width=0.48\textwidth]{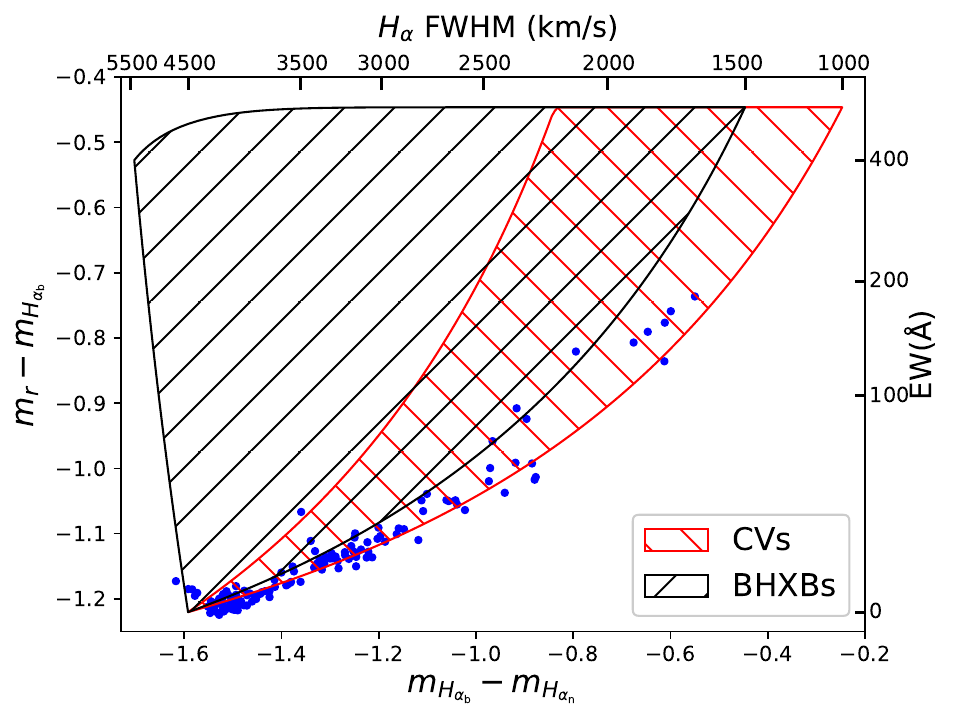}
    \caption{Theoretical distributions of BHXBs (area covered with black lines) and CVs (area covered with red lines) in $H_{\alpha}$ color--color diagram, using idealized filters centered at 6564.61 \AA \ and effective widths of 30 \AA\ ( $H_{\alpha_{\rm n}}$ filter), 130 \AA\ ($H_{\alpha_{\rm b}}$ filter), and 400 \AA\ ($r$-band filter). The blue dots are the CVs in the LRS sample with color corrections.
    \label{theo_dis_ha.fig}}
\end{figure}

Figure \ref{theo_dis_ha.fig} shows the theoretical distributions of BHXBs (black lines) and CVs (red lines) in the $H_{\alpha}$ color--color diagram, which is consistent with the observed distribution of CVs (blue points).
Notably, the theoretical photometric fluxes were calculated under the assumption that the $H_{\alpha}$ profiles are symmetric, while most of the observed $H_{\alpha}$ profiles in our sample were asymmetric. 
To address this, we applied additional corrections to the asymmetric $H_{\alpha}$ profiles by adjusting the observed colors of CVs to match the theoretical colors (Appendix \ref{correction_colours.sec}).

\section{Orbital parameter estimation}
\label{fwhm_ew_k2.sec}

\subsection{The FWHM-$K_{\rm 2}$ correlation}
\label{FWHM_K2.sec}

The FWHM can be used to estimate the semi-amplitude of radial velocity of the donor star in compact binaries.
As shown in \citet{2015ApJ...808...80C}, for a system with $H_{\alpha}$ emission line dominated by the accretion disk, the relationship between the FWHM and $K_{\rm 2}$ is
\begin{equation} \label{eq3}
\frac{K_{\rm 2}}{\rm{FWHM}} = \frac{\sqrt{\alpha f(q)}}{2},
\end{equation}
where 
\begin{equation} \label{eq4}
f(q) = \frac{0.49(1+q)^{-1}}{0.6+q^{2/3} \rm{ln} (1+q^{-1/3})}.
\end{equation}
Here $q$ ($=M_2/M_1$) is the mass ratio and $\alpha$ ($< 1$) is the ratio of the characteristic disk radius to the Roche lobe of the accreting compact object.
The subscripts ``1" and ``2" represent the accretor and donor, respectively.
For BHXBs, the $\sqrt{f(q)}$ values vary slightly ($\simeq$ 0.69--0.77) since their mass ratios do not vary much ($\simeq$ 0.05--0.15).
Consequently, \citet{2015ApJ...808...80C} derived approximately linear correlations between $K_{\rm 2}$ and FWHM, as $K_{\rm 2} = 0.233(13)\ \rm{FWHM}$ for BXBHs and $K_{\rm 2} = 0.169(16)\ \rm{FWHM}$ for long-period CVs.

\begin{figure}
    \center
    \includegraphics[width=0.48\textwidth]{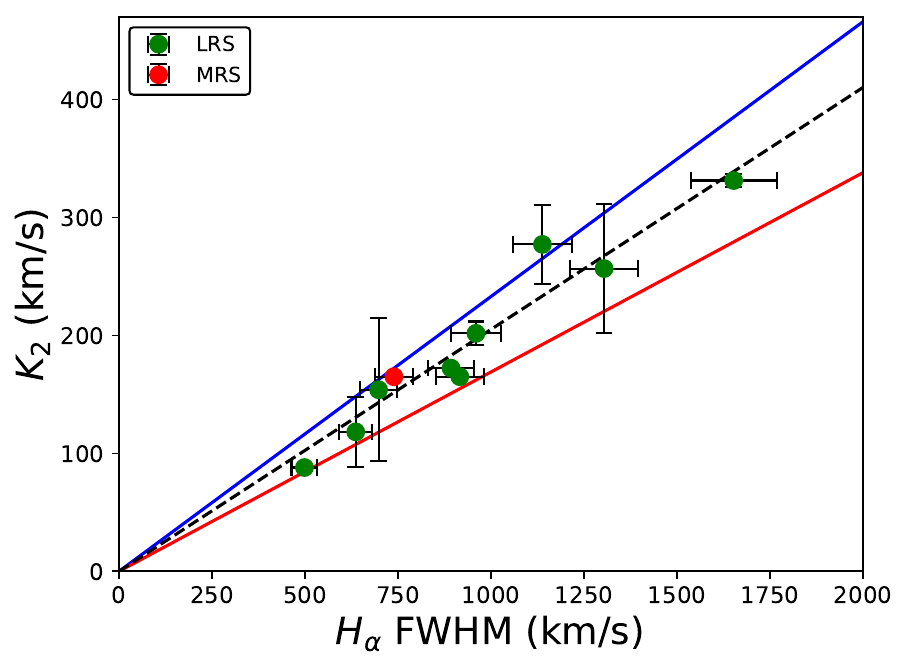}
    \caption{The correlation between the FWHM and the $K_{\rm 2}$. The green dots are the results from the LRS sample, while the red dot marks the result from the MRS sample. The blue and red lines are the best fitting in previous studies \citep{2015ApJ...808...80C} for BHXBs and CVs, and the black dashed line is the best fitting for our CVs sample in this paper.}
    \label{fwhm_Kv.fig}
\end{figure}

\begin{figure}
    \center
    \includegraphics[width=0.48\textwidth]{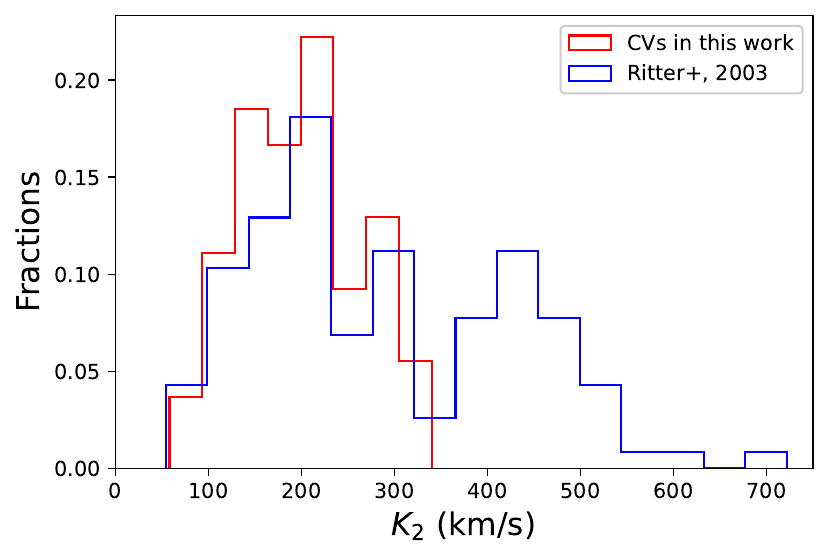}
    \caption{Comparison of the $K_{\rm 2}$ distribution between the CVs in our sample and those in \cite{2003A&A...404..301R}.}
    \label{k2_dis.fig}
\end{figure}

For CVs in our sample, we cross-matched them with previous studies \citep{1974ApJ...193..191R,1980MNRAS.192..127B,1984ApJ...286..747H,1985ApJ...289..300P,2000MNRAS.318....9B,2000MNRAS.313..383N,2002MNRAS.329..597M,2002MNRAS.337.1215N,2003A&A...404..301R,2009MNRAS.399.1534C,2010MNRAS.402.1824C,2019A&A...622A.186S} to collect $K_{\rm 2}$ and $q$ values.
Ten systems (nine in the LRS sample and one in the MRS sample) have $K_{\rm 2}$ and $q$ measurements (Table \ref{fwhm_K2.tab}).
Using these systems, we derived a similar linear correlation of $K_{\rm 2} = 0.205(18)\ \rm{FWHM}$ (Figure \ref{fwhm_Kv.fig}).  
Applying this correlation, we calculated the $K_{\rm 2}$ values for all the CVs in our sample (Table \ref{K2.tab}).
Additionally, we calculated the $K_{\rm 2}$ values for the CVs in \cite{2003A&A...404..301R} based on their orbital parameters (i.e., mass ratio, inclination, white dwarf mass, and orbital period) following
\begin{equation}
    \frac{M_{1} \, \textrm{sin}^3 i} {(1+q)^{2}} = \frac{P \, K_{2}^{3}}{2\pi G}.
\label{mf}
\end{equation}
Figure \ref{k2_dis.fig} compares the $K_{\rm 2}$ distribution of CVs from our sample and those in \cite{2003A&A...404..301R}.
To avoid contamination from other types of objects, we excluded CVs with multiple classifications from SIMBAD (Section \ref{types.sec}).
The $K_{\rm 2}$ distribution of CVs from \cite{2003A&A...404..301R} shows two populations, separated at $K_{\rm 2} \approx$ 350 km/s, corresponding to CVs with orbital periods below and above the period gap (i.e., 2--3 hours), respectively. 
Compared to \cite{2003A&A...404..301R}, our CV sample shows an absence of CVs in the secondary population with $K_{\rm 2} \gtrsim$ 350 km/s, which can be attributed to two factors.
First, as shown by \citep{2018MNRAS.473.5195C}, approximately 99\% of CVs exhibit FWHM values exceeding 1000 km/s, while only about 4\% have FWHM values beyond 2000 km/s.
Given this small fraction, the limited size of our CV sample makes it unlikely to identify CVs with FWHM$>$2000 km/s.
Second, CVs exhibit a wider range of mass ratios compared to BHXBs, leading to significant variability in $\sqrt{f(q)}$ values ($\simeq$ 0.44--0.72 for $q$ from 0.1 to 1). 
As shown by \cite{2015ApJ...808...80C}, for CVs with $q>$0.2 and FWHM$<$1800 km/s, the relationship between $K_{\rm 2}$ and FWHM follows a tight linear correlation with small scatter, while for CVs with $q<$0.2 and FWHM$>$1800 km/s, the ratio between $K_{\rm 2}$ and FWHM increases significantly, accompanied by larger scatter.
Consequently, for CVs below the period gap, the $K_{\rm 2}$ values calculated using $K_{\rm 2} = 0.205(18)\ \rm{FWHM}$ may be underestimated.
This indicates for CVs in our sample with FWHM values larger than 1400 km/s ($\approx$9 systems), their $K_{\rm 2}$ values are likely underestimated. 
Thus, we recommend first determining the mass ratio from light curves \citep{2009MNRAS.398.2110W} and then calculating $K_2$ using Equations \ref{eq3} and \ref{eq4} for more accurate results.

\begin{table*}
\caption{Database of cataclysmic variables in our samples. \label{fwhm_K2.tab}}
\centering
\setlength{\tabcolsep}{5mm}
\begin{center}
\begin{tabular}{lccccc}
\hline\noalign{\smallskip}
Name & FWHM$^{b}$ & $K_{\rm 2}^{a}$ & $q^{a}$ & Type & References\\
 & (km/s) & (km/s) & & \\
\hline\noalign{\smallskip}
UU Aqr & $1150.23\pm80.52$ & $277.13\pm33.18$ & $0.30\pm0.07$ & LRS & (1) \\
GK Per & $657.68\pm46.04$ & $118.20\pm29.39$ & $0.55\pm0.21$ & LRS & (1), (2) \\
UX UMa & $1314.42\pm92.01$ & $256.67\pm54.49$ & $0.43\pm0.06$ & LRS & (1) \\
EM Cyg & $973.80\pm68.17$ & $201.78\pm10.02$ & $0.80\pm0.04$ & LRS & (1), (7), (8) \\
T Aur & $907.83\pm63.55$ & $172.28\pm1.86$ & $0.90\pm0.20$ & LRS & (1), (10), (11) \\
BG CMi & $717.31\pm50.21$ & $153.94\pm60.61$ & 0.38-0.63 & LRS & (1), (12) \\
SS Cyg & $931.00\pm65.17$ & $165.00\pm1.00$ & $0.69\pm0.03$ & LRS & (3), (4) \\
IP Peg & $1660.52\pm116.24$ & $331.30\pm5.80$ & $0.48\pm0.01$ & LRS & (5), (6) \\
SY Cnc & $525.49\pm36.78$ & $88.00\pm2.90$ & $1.18\pm0.14$ & LRS & (9) \\
SS Cyg & $739.93\pm51.80$ & $165.00\pm1.00$ & $0.69\pm0.03$ & MRS & (3), (4) \\
\noalign{\smallskip}\hline
\end{tabular}
\end{center}
NOTE. $^a$ means that this parameter was derived from references; $^b$ means that this parameter was estimated from $H_{\alpha}$ line in this work. References including: (1) \cite{2003A&A...404..301R}, (2) \cite{2002MNRAS.329..597M}, (3) \cite{2002MNRAS.337.1215N}, (4) \cite{1984ApJ...286..747H}, (5) \cite{2010MNRAS.402.1824C}, (6) \cite{2000MNRAS.318....9B}, (7) \cite{1974ApJ...193..191R}, (8) \cite{2000MNRAS.313..383N}, (9) \cite{2009MNRAS.399.1534C}, (10) \cite{1980MNRAS.192..127B}, (11) \cite{2019A&A...622A.186S}, (12) \cite{1985ApJ...289..300P}.
\end{table*}

\begin{table*}
\caption{Calculated $K_{\rm 2}$ values of CVs in our sample. \label{K2.tab}}
\centering
\begin{center}
\begin{tabular}{cccccc}
\hline\noalign{\smallskip}
Name & R.A. & Decl. &  FWHM & $K_{\rm 2}$ & Database\\
 & (deg) & (deg) & (km/s) & (km/s)  \\
\hline\noalign{\smallskip}
V482 And & 5.14420 & 40.41837 & $260\pm18$ & $53\pm4$ & LRS \\
FBS 0019+348 & 5.45187 & 35.08089 & $781\pm55$ & $160\pm11$ & LRS \\
ATO J005.9853+42.7854 & 5.98529 & 42.78550 & $607\pm43$ & $125\pm9$ & LRS \\
HQ And & 7.89952 & 43.81814 & $559\pm39$ & $115\pm8$ & LRS \\
IW And & 15.28711 & 43.39043 & $282\pm20$ & $58\pm4$ & LRS \\
GN Psc & 16.27670 & 9.58562 & $101\pm7$ & $21\pm1$ & MRS \\
ATO J016.9284+48.7552 & 16.92846 & 48.75520 & $1630\pm114$ & $334\pm23$ & LRS \\
2MASS J01275053+3808120 & 21.96081 & 38.13663 & $1318\pm92$ & $270\pm19$ & LRS \\
TW Tri & 24.15423 & 32.01113 & $676\pm47$ & $139\pm10$ & LRS \\
AY Psc & 24.23109 & 7.27465 & $1070\pm75$ & $219\pm15$ & LRS \\
SDSS J013855.86+242939.2 & 24.73278 & 24.49423 & $1175\pm82$ & $241\pm17$ & LRS \\
LAMOST J015016.17+375618.9 & 27.56740 & 37.93861 & $987\pm69$ & $202\pm14$ & LRS \\
AI Tri & 30.95255 & 29.99047 & $1406\pm98$ & $288\pm20$ & LRS \\
CRTS J021110.2+171624 & 32.79252 & 17.27336 & $946\pm66$ & $194\pm14$ & LRS \\
HS 0220+0603 & 35.75695 & 6.28044 & $1476\pm103$ & $303\pm21$ & LRS \\
V573 And & 36.21859 & 42.44824 & $233\pm16$ & $48\pm3$ & LRS \\
WY Tri & 36.25193 & 32.99878 & $1659\pm116$ & $340\pm24$ & LRS \\
LAMOST J030531.06+231815.0 & 46.37946 & 23.30418 & $540\pm38$ & $111\pm8$ & LRS \\
CRTS J032034.5+203607 & 50.14383 & 20.60200 & $1079\pm76$ & $221\pm15$ & LRS \\
GK Per & 52.80001 & 43.90423 & $658\pm46$ & $135\pm9$ & LRS \\
\noalign{\smallskip}\hline
\end{tabular}
\end{center}
NOTE. This table is available in its entirety in machine-readable and Virtual Observatory (VO) forms in the online journal. A portion is shown here for guidance regarding its form and content.
\end{table*}

\subsection{Systematic parameters}
\label{orbit.sec}

The double-peaked $H_{\alpha}$ emission lines, originating from the accretion disk, provide a way to estimate systematic parameters (e.g., mass ratio $q$, inclination angle $i$, and the mass of the accretor $M_{1}$) for CVs and XBs.
Three relations have been given between the systematic parameters and $H_{\alpha}$ profile parameters as follows \citep{2016ApJ...822...99C,2022MNRAS.516.2023C}:
\begin{equation}\label{eq15}
\frac{\rm DP}{\rm FWHM} = 3^{1/3}(1+q)^{2/3}\beta\sqrt{\alpha f(q)},
\end{equation}
\begin{equation}\label{eq16}
i = 93.5T+23.7,    \ \ \     {\rm with} \ T=\frac{h-f_{T}}{h},
\end{equation}
and
\begin{equation}\label{eq17}
M_{1}=3.45\times10^{-8}P_{\rm obs}[(0.63W+145)/\beta]^{3} \ M_{\odot}.
\end{equation}
Here DP is the double peak separation, $h$ is the height of the $H_{\alpha}$ line, $f_{T}$ is the flux at the central depression or trough (between the peaks), $T$ is the dimensionless depth of the trough (i.e. normalized to the double peak height), and $W$ is the full-width at half-maximum of each Gaussian function.
The parameter $\beta$ serves as a correction factor, accounting for that the outer disc material is sub-Keplerian.
Previous studies \citep[e.g.,][]{1988ApJ...324..411W,2016ApJ...822...99C,2022MNRAS.516.2023C} yielded a $\beta$ of $\approx$0.84.

To verify the precision of these parameters estimated by this method, we cross-matched our CVs sample with the catalog from \cite{2003A&A...404..301R}, which provided extensive information (e.g., orbital period, mass ratio, inclination angle, $M_{1}$ and $M_{2}$) for CVs and LMXBs.
We visually checked the spectra of the common sources and then selected six CVs with double-peaked $H_{\beta}$ or $H_{\alpha}$ lines to estimate their systematic parameters.
Unfortunately, all six systems have only one LRS observation at a random orbital phase, leading to an asymmetric structure in their $H_{\beta}$ or $H_{\alpha}$ lines.
We used two fitting models to estimate their systematic parameters (Appendix \ref{est_paras.sec}). 
Figure \ref{Ha_Hb.fig} shows the results using those fitting models. 
The troughs of the $H_{\alpha}$ lines are shallower compared to those of the $H_{\beta}$ lines, since the emitting region of the $H_{\alpha}$ line is located further out than that of the $H_{\beta}$ line.
This results in a smaller value and larger uncertainty of DP of the $H_{\alpha}$ line.
Consequently, the systematic parameters derived from the $H_{\beta}$ line are more accurate than those derived from the $H_{\alpha}$ line.

\begin{figure*}
    \center
    \includegraphics[width=1\textwidth]{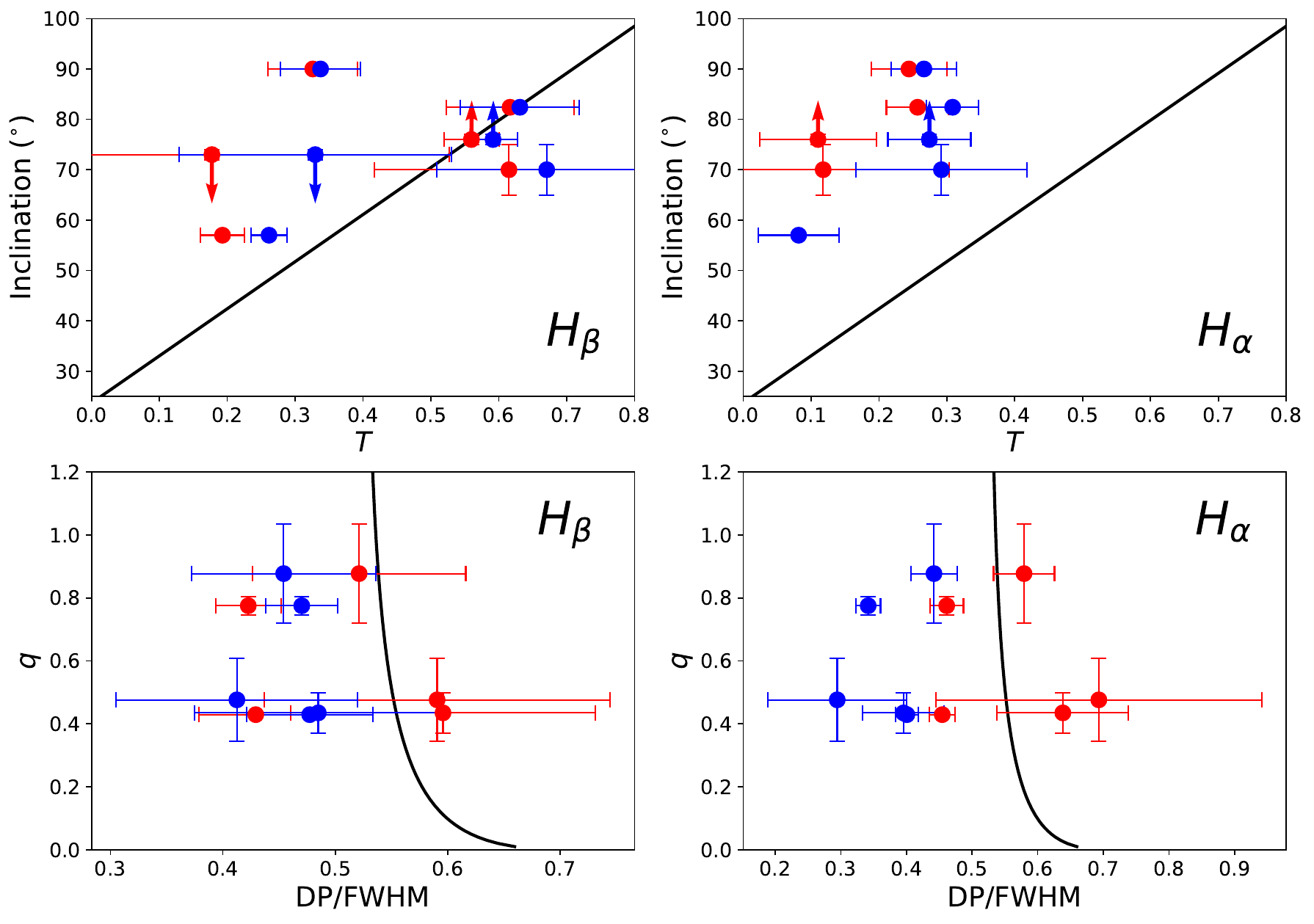}
    \caption{Top panel: Comparison of the $H_{\beta}$ and $H_{\alpha}$ line parameter $T$ with the inclination angle for CVs in our sample. The solid lines represent the results from Equation \ref{eq16}. The red and blue dots represent the $T$ values estimated from the symmetric Gaussian fitting and the three single Gaussian fitting, respectively.
    Bottom panel: Comparison of the $H_{\beta}$ and $H_{\alpha}$ line parameter DP/FWHM with the mass ratio $q$ for CVs in our sample. The solid lines represent the results from Equation \ref{eq15} with $\alpha=0.42$ and $\beta=0.84$. The red and blue dots represent the DP/FWHM values estimated from the symmetric Gaussian fitting and the three single Gaussian fitting, respectively.
    }
    \label{massr_incl.fig}
\end{figure*}

Figure \ref{massr_incl.fig} compares the system parameters derived from Equations. \ref{eq15} and \ref{eq16} with those from \cite{2003A&A...404..301R}. 
On the one hand, the systematic parameters obtained from the $H_{\beta}$ lines are closer to the values reported from \cite{2003A&A...404..301R}, compared to those derived from the $H_{\alpha}$ lines. 
On the other hand, the deviation between the parameters calculated from Equations. \ref{eq15} and \ref{eq16} and those from \cite{2003A&A...404..301R} are mainly due to the limited number of spectroscopic observations and the low spectral resolution.
The weak correlation between our results and those from \cite{2003A&A...404..301R} suggest that more observations covering various orbital phases are required to eliminate the effects of line asymmetry and improve the accuracy of systematic parameter estimations.

\section{Summary}
\label{summary.sec}

We used the LAMOST LRS DR9 emission catalog to investigate the methods proposed by \citet{2015ApJ...808...80C,2016ApJ...822...99C,2018MNRAS.473.5195C}, \citet{2018MNRAS.481.4372C}, and \citet{2022MNRAS.516.2023C} for measuring systematic parameters and searching for compact objects.

First, the distributions of FWHM and EW provide a way for rapid stellar classification for objects with different $H_{\alpha}$ emission profiles.
Our sample includes a significant number of YSOs, Be stars, and CVs.
Using LAMOST LRS and MRS data (Section \ref{FWHM_EW.sec}), we found that the YSOs can be divided into two subpopulations: a more active sample ($H_{\alpha}$ EW $\gtrsim$ 3 \AA) and a less active sample (EW $\lesssim$ 3 \AA).
The more active YSOs and Be stars are located in a similar region in the FWHM--EW plane.
CVs display a wide range of FWHM and EW values, and the CVs with broad $H_{\alpha}$ lines (i.e., FWHM $\gtrsim$ 1000 km/s) can be easily distinguished in the FWHM--EW plane.
In addition to spectroscopic methods, the $H_{\alpha}$ color--color diagram can also be used for stellar classifications (Section \ref{color_disgram.sec}).
The distribution of CVs in the $H_{\alpha}$ color--color diagram is clearly different from that of YSOs and Be stars, due to the differences in their continuum (indicated by $m_{r}$ - $m_{H_{\alpha_{\rm b}}}$) and $H_{\alpha}$ profiles (indicated by $m_{H_{\alpha_{\rm b}}}$ - $m_{H_{\alpha_{\rm n}}}$).
We also attempted to find a new set of idealized filters to select CVs and BHXBs with high significance.
Consequently, we derived optimal effective widths of 30 \AA, 130 \AA, and 400 \AA\ for the $H_{{\alpha}_{\rm n}}$, $H_{{\alpha}_{\rm b}}$, and $r$-band filters, respectively.

Second, we used the FWHM and EW values estimated from $H_{\alpha}$ and $H_{\beta}$ emission lines to calculate systematic parameters of CVs in our sample.
According to the FWHM values and the $K_{2}$ results from previous studies, we obtained a new linear correlation for CVs, $K_{\rm 2} = 0.205(18)\ \rm{FWHM}$, which is similar to the correlation derived by \cite{2015ApJ...808...80C} (Section \ref{FWHM_K2.sec}).
Systematic parameters ($q$, $i$, and $M_{1}$) can be derived from the $H_{\beta}$ or $H_{\alpha}$ profile for BHXBs and CVs using the following correlations (Section \ref{orbit.sec}):
\begin{itemize}
\item[$\centerdot$] $\frac{\rm DP}{\rm FWHM} = 3^{1/3}(1+q)^{2/3}\beta\sqrt{\alpha f(q)}$;\\
\item[$\centerdot$] $i = 93.5T+23.7$;\\
\item[$\centerdot$] $M_{1}=3.45\times10^{-8}P_{\rm obs}[(0.63W+145)/\beta]^{3} \ M_{\odot}$.
\end{itemize}
We estimated the systematic parameters for six CVs in our sample with one LRS observation.
The final results suggest that systematic parameters from one (low-resolution) spectroscopic observation have significant uncertainties due to the asymmetrical structure of their $H_{\beta}$ or $H_{\alpha}$ lines.
Sufficient orbital coverage of the observed spectra is crucial for accurately estimating systematic parameters.

These methods open up a new path to search for BHs in accretion states and calculate their systematic parameters.
With more large spectroscopic and photometric surveys, more compact binary candidates are expected to be discovered, which are key to understanding binary formation and evolution.

\begin{acknowledgements}

We thank the anonymous referee for helpful comments and suggestions that have significantly improved the paper. 
We thank Yongkang Sun for a very useful discussion about cataclysmic variables.
The Guoshoujing Telescope (the Large Sky Area Multi-Object Fiber Spectroscopic Telescope LAMOST) is a National Major Scientific Project built by the Chinese Academy of Sciences. Funding for the project has been provided by the National Development and Reform Commission. LAMOST is operated and managed by the National Astronomical Observatories, Chinese Academy of Sciences. 
%This work uses data obtained through the Telescope Access Program (TAP), which has been funded by the TAP member institutes. 
%This work presents results from the European Space Agency (ESA) space mission {\it Gaia}. {\it Gaia} data are being processed by the {\it Gaia} Data Processing and Analysis Consortium (DPAC). Funding for the DPAC is provided by national institutions, in particular the institutions participating in the {\it Gaia} MultiLateral Agreement (MLA). The {\it Gaia} mission website is https://www.cosmos.esa.int/gaia. The {\it Gaia} archive website is https://archives.esac.esa.int/gaia. 
This research has made use of the SIMBAD database, operated at CDS, Strasbourg, France, and of the VizieR catalogue access tool, CDS, Strasbourg, France.
This work made use of Astropy\footnote{http://www.astropy.org}, a community-developed core Python package and an ecosystem of tools and resources for astronomy.
%This research made use of Photutils (Bradley et al. 2020), an Astropy package for detection and photometry of astronomical sources. 
This work was supported by National Key Research and Development Program of China (NKRDPC) under grant No. 2023YFA1607901, 
Strategic Priority Program of the Chinese Academy of Sciences under grant No. XDB1160302,
National Science Foundation of China (NSFC) under grant Nos. 11988101/12273057/11833002/12090042, and Chinese Academy of Sciences Basic Frontier Science Research Program from 0 to 1 Original Innovation Project under grant No. ZDBS-LY-SLH002.

\end{acknowledgements}

\bibliographystyle{aasjournal}
\bibliography{main.bib}{}   

 \clearpage
\appendix
\renewcommand*\thetable{\Alph{section}.\arabic{table}}
\renewcommand*\thefigure{\Alph{section}\arabic{figure}}

\section{Comparison between observed and theoretical FWHMs for single-Gaussian profiles}
\label{fwhmm_fwhmg_single_profiles.sec}

\setcounter{figure}{0}

\begin{figure}
    \center
    \includegraphics[width=0.48\textwidth]{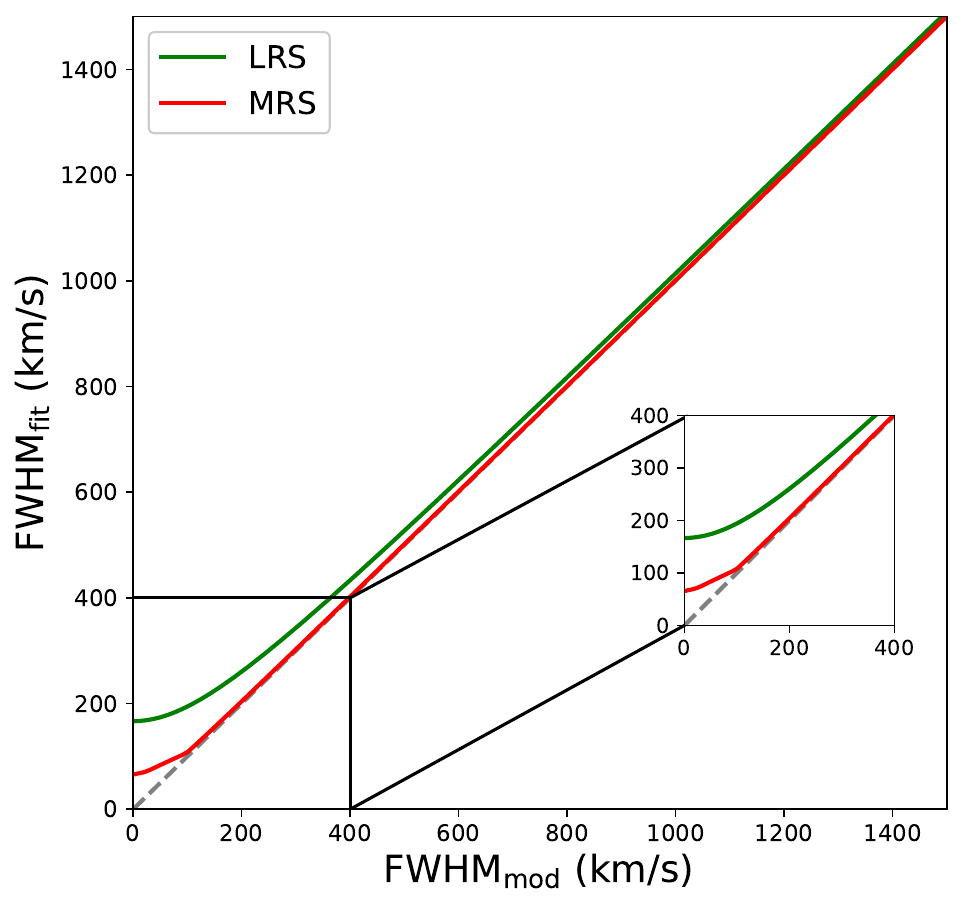}
    \caption{Comparison between FWHM$_{\rm mod}$ and FWHM$_{\rm fit}$. The green line and red line represent the results derived from the Gaussian profiles with the resolution of LRS ($\sim$ 1800) and MRS ($\sim$ 7500), respectively. The FWHM$_{\rm mod}$ is used to generate the theoretical single-Gaussian profile and FWHM$_{\rm fit}$ is derived by fitting a single-Gaussian function to this profile.}
    \label{fwhm_correction.fig}
\end{figure}

In Section \ref{data.sec}, we measured the FWHM of the $H_{\alpha}$ using a Gaussian function for both LAMOST LRS and MRS data. However, FWHM values depend on spectral resolution. Therefore, to compare the FWHM values derived from spectra in different surveys, the resolution effects should first be corrected.
We synthesized a group of theoretical single-Gaussian profiles with different FWHM$_{\rm mod}$ and resampled these profiles to match the resolutions of LRS ($\sim$1800) and MRS ($\sim$7500).
FWHM$_{\rm fit}$ is derived from a single Gaussian fitting to the spectra with low and medium resolutions.
The comparisons between FWHM$_{\rm mod}$ and FWHM$_{\rm fit}$ for different resolutions of LRS and MRS are shown in Figure \ref{fwhm_correction.fig}.
We used these relations to correct the FWHM values derived from LAMOST observations to modeled values for analysis in this paper.

\section{Comparison between observed and theoretical FWHMs for double-peaked profiles}
\label{fwhmm_fwhmg.sec}

\setcounter{table}{0}
\setcounter{figure}{0}

In Section \ref{color_disgram.sec}, to include the FWHM and EW information in the $H_{\alpha}$ color--color diagram, we need to synthesize a set of double-peaked $H_{\alpha}$ profiles with varying FWHM and EW values and calculate their photometry in the three bands.
However, the observed data in the $H_{\alpha}$ color--color diagram were derived from the LAMOST LRS data.
Therefore, the synthesized spectra also need to match the LAMOST low resolution.

\begin{figure}[h]
    \center
    \includegraphics[width=0.48\textwidth]{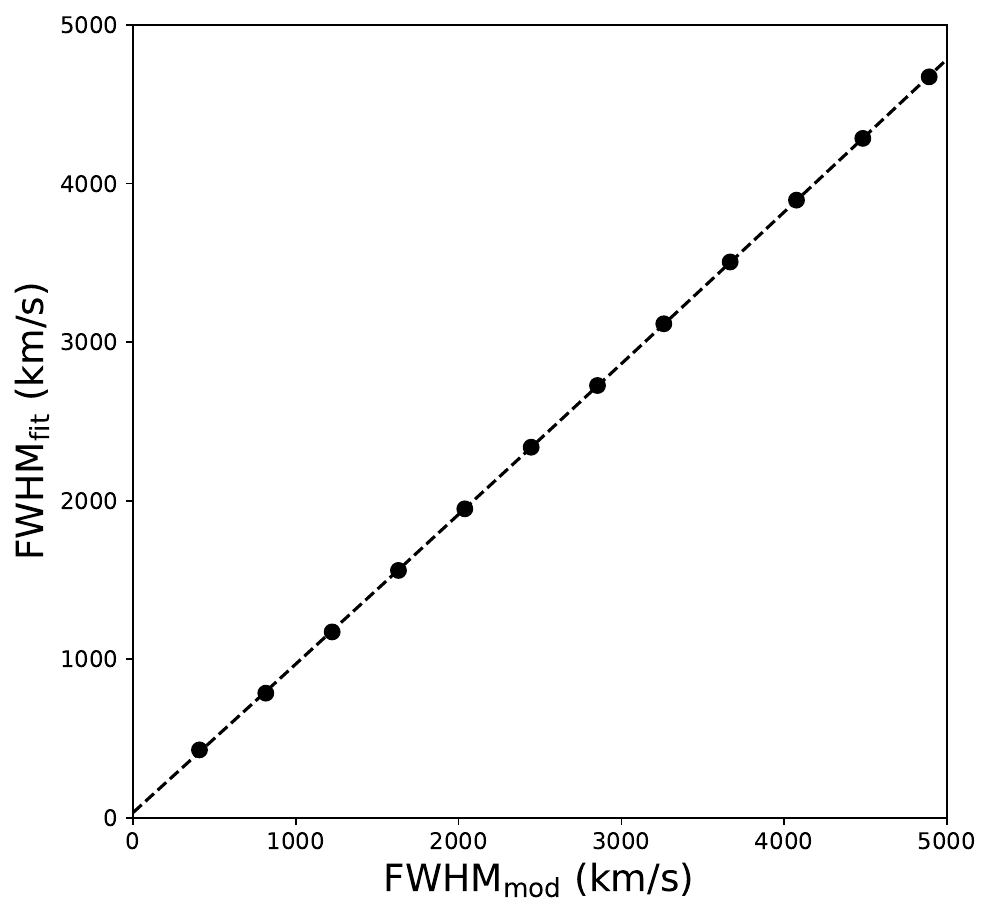}
    \caption{The FWHM$_{\rm fit}$ from Gaussian fit compared to the model FWHM$_{\rm mod}$. The black dashed line is the best fitting using a quadratic fit.
    The FWHM$_{\rm mod}$ is used to generate the theoretical symmetrically double-peaked profile and FWHM$_{\rm fit}$ is derived by fitting a single-Gaussian function to this profile.
    }
    \label{fwhm_fwhmg.fig}
\end{figure}

To derive the correction, we generated a set of double-peaked $H_{\alpha}$ profiles with different FWHM$_{\rm mod}$ values (Equation \ref{eq6}), and then resampled these spectra at the LAMOST low resolution.
We used a single Gaussian function to fit these spectra to derive the FWHM$_{\rm fit}$ values, which can be considered as FWHM values from observed spectra.
The correlation between FWHM$_{\rm mod}$ and FWHM$_{\rm fit}$ was obtained from a quadratic fit, as shown in Figure \ref{fwhm_fwhmg.fig}:
\begin{equation} \label{eq8}
{\rm FWHM}_{\rm fit} = 3.32\times10^{-6}{\rm FWHM}_{\rm mod}^{2} + 0.93349{\rm FWHM}_{\rm mod} + 33.12
\end{equation}
Finally, we used Equation \ref{eq8} to convert the theoretical FWHM values to observed values to include FWHM in the $H_{\alpha}$ color--color diagram.

\section{Corrections to the observed $H_{\alpha}$ colors of the CV systems}
\label{correction_colours.sec}

\setcounter{figure}{0}

\begin{figure*}
    \center
    \includegraphics[width=0.98\textwidth]{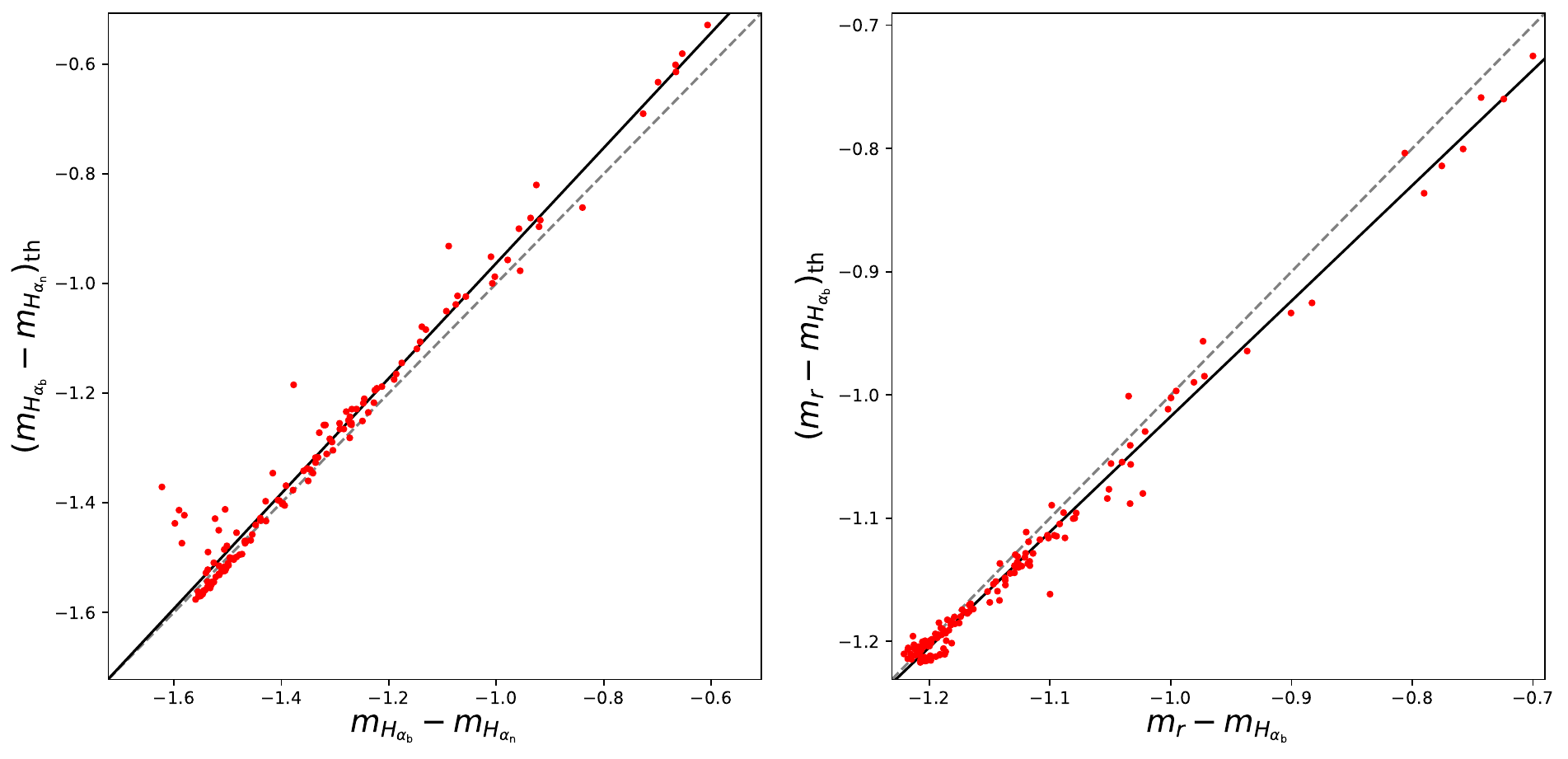}
    \caption{The comparisons between observed and theoretical colors. The black lines are the best fittings using linear fitting. The values of the theoretical colors were derived using the FWHM and EW values of CVs with FWHM $>$ 1000 km/s in the LRS sample.}
    \label{correction_colour.fig}
\end{figure*}

In Section \ref{colour_dis.sec}, we studied the theoretical distributions of BHXBs and CVs.
In Figure \ref{theo_dis_ha.fig}, the black and red lines were derived by constructing a set of symmetric double-peaked $H_{\alpha}$ profiles, reducing their resolution to the LAMOST/LRS level, and fitting the spectra with a single Gaussian function, with the correction method from Appendix \ref{fwhmm_fwhmg.sec}.
Therefore, for the blue points in Figure \ref{theo_dis_ha.fig}, which are CV systems, the non-equally double-peaked effect of the $H_{\alpha}$ lines should be corrected for accurate comparison with these lines.

Here we did the corrections to observed magnitudes $m_{\rm H_{\alpha _{n}}}$ and $m_{\rm H_{\alpha _{b}}}$, $m_{\rm r}$ for the CV systems.
First, using the optimal set of filters with effective widths of 30 \AA, 130 \AA, and 400 \AA, we calculated the observed magnitudes and colors for the CVs in our LRS sample.
Second, we estimated the FWHM$_{\rm fit}$ and EW values of observed $H_{\alpha}$ profiles using a Gaussian function, and then converted FWHM$_{\rm fit}$ values to theoretical values FWHM$_{\rm mod}$ using Equation \ref{eq8} in Appendix \ref{fwhmm_fwhmg.sec}.
The values of the theoretical colors were obtained from Equations \ref{eq11}, \ref{eq12} using the values of EW and DP($=$FWHM$_{\rm mod}/$1.785).
Figure \ref{correction_colour.fig} shows two linear fittings between the observed  and theoretical colors as follows,
\begin{equation}\label{eq13}
(m_{H_{\alpha_{\rm b}}} - m_{H_{\alpha_{\rm n}}})_{\rm th} = 1.0510 \times (m_{H_{\alpha_{\rm b}}} - m_{H_{\alpha_{\rm n}}}) + 0.08784
\end{equation}
and
\begin{equation}\label{eq14}
(m_{r} - m_{H_{\alpha_{\rm b}}})_{\rm th} = 0.9375 \times (m_{r} - m_{H_{\alpha_{\rm b}}}) - 0.07995
\end{equation}
A significant offset can be seen between $(m_{H_{\alpha_{\rm b}}} - m_{H_{\alpha_{\rm n}}})_{\rm th}$ and $m_{H_{\alpha_{\rm b}}} - m_{H_{\alpha_{\rm n}}}$, indicating that most of the observed $H_{\alpha}$ profiles in our sample exhibit non-equal double-peaked profiles.
These two relations (Equations \ref{eq13}, \ref{eq14}) were used to correct the observed colors of CVs for plotting them in the theoretical $H_{\alpha}$ color--color diagram.

\section{Estimation of systematic orbital parameters for the six CVs}
\label{est_paras.sec}

\setcounter{figure}{0}

We calculated systematic orbital parameters for six CVs: LX Ser, J0107+4845 (hereafter J0107), J0644+3344 (hereafter J0644), GK Per, UX UMa, and T Aur.
We used two models to fit the $H_{\beta}$ and $H_{\alpha}$ lines for the six CVs: (1) a symmetric double-Gaussian function, with the wavelength ranges for the fitting being 4855 \AA \ to 4870 \AA \ for $H_{\beta}$ and 6550 \AA \ to 6575 \AA \ for $H_{\alpha}$; (2) three single-Gaussian functions, which are used to fitting the two peaks and the central trough, respectively.
We note that in most of our sources, the two peaks of the Balmer lines are asymmetrical due to limited observations at random orbital phases. In contrast, \citet{2022MNRAS.516.2023C} performed many more observations to cover the whole phase, thus deriving averaged Balmer lines with consistently high double peaks.

To investigate the rationality of the second model, we created an asymmetrical emission line profile, by adding a hot spot to a double-peaked line, which can be expressed as follows \citep{1994ApJ...436..848O}:
\begin{equation}\label{eq18}
\begin{split}
F(u) \propto \int_{r_{1}}^{r_{z}} \frac{r^{3/2-\gamma}dr}{(1-u^{2}r)^{1/2}}(1-\omega \ {\rm sin}\psi \ ur^{1/2}) \\
\times [1+(2u \ {\rm sin}i \ {\rm tan}i)^{2}r(1-u^{2}r)]^{1/2},
\end{split}
\end{equation}
where $u$ is the dimensionless radial velocity, that is the ratio of the RV at the outer edge of the disk to that at a radius $r$. 
$r_{1}$ is the ratio of the inner radius to the outer radius of the disk, and $r_{z}={\rm min}(1, u^{-2})$ \citep{1981AcA....31..395S}. 
$\gamma$ is the exponent of the power-law emissivity law that describes the accretion disc \citep{1973A&A....24..337S,1981AcA....31..395S}.
$\omega$ represents the flux variation amplitude of the lines due to the hot spot.
For symmetric double-peaked lines ($\psi=0$), the height $h$ can be expressed by $F(-1)$ or $F(1)$; while the height $h$ for any $\psi$ can be represent by $\frac{F(-1)+F(1)}{2}$.
Therefore, it's feasible to use the averaged height of the double peaks as the height of $H_{\alpha}$ in the second model.

\begin{figure*}
    \center
    \includegraphics[width=1\textwidth]{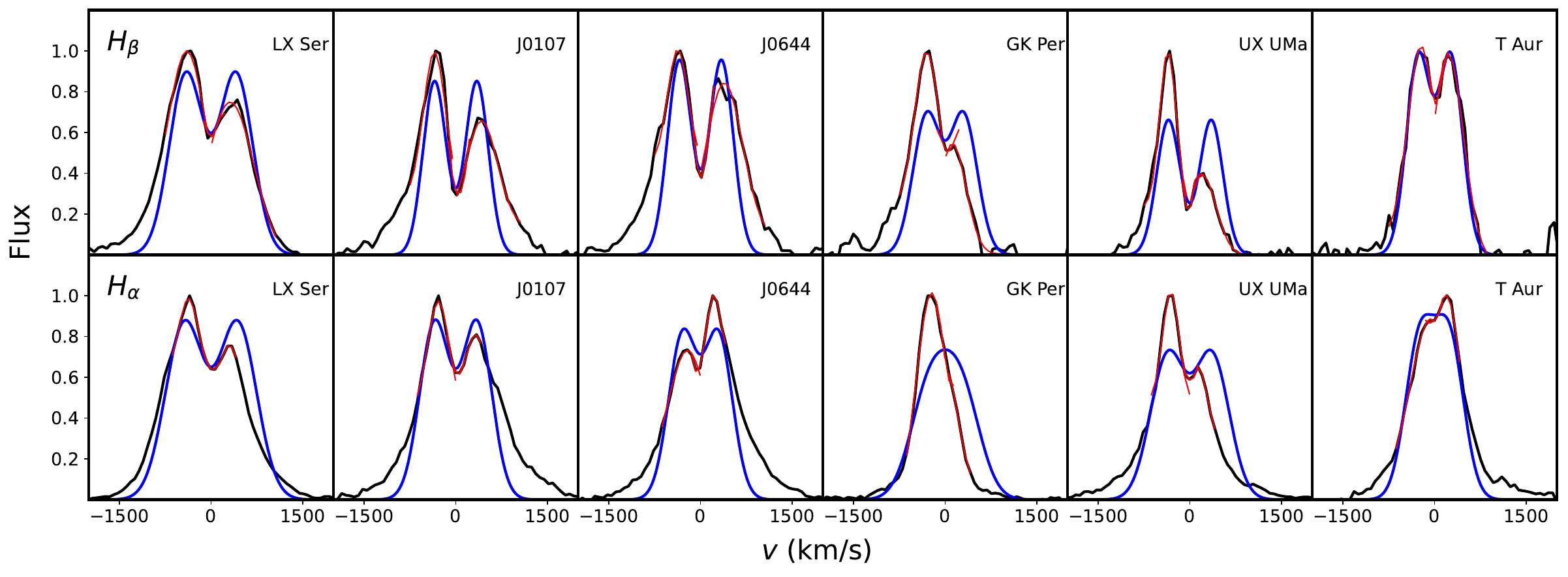}
    \caption{Profile fitting for $H_{\beta}$ and $H_{\alpha}$ lines of the CVs in our sample. The black lines are the observed $H_{\beta}$ and $H_{\alpha}$ profiles. The blue lines are the best fitting from the symmetric Gaussian model, while the red lines indicate the results of three single Gaussian fits.}
    \label{Ha_Hb.fig}
\end{figure*}

As shown in Figure \ref{Ha_Hb.fig}, the troughs of the $H_{\beta}$ lines are significantly deeper than that of the $H_{\alpha}$ lines. 
This is expected since the emitting region of the $H_{\alpha}$ line is located further out than that of the $H_{\beta}$ line, leading to slower rotation and thus closer peaks of the $H_{\alpha}$ line.
This results in a smaller value and larger uncertainty of DP of the $H_{\alpha}$ line.
Consequently, the systematic parameters derived from the $H_{\beta}$ line are more accurate than those derived from the $H_{\alpha}$ line.
It's interesting that the $H_{\alpha}$ and $H_{\beta}$ lines observed in J0644 show opposite profiles, indicating that the $H_{\alpha}$ line may be contaminated (e.g., by the magnetic activity of the companion).

\end{document}